\documentclass{agujournal2019}
\usepackage{url} 
\usepackage{natbib}		     	        
\usepackage{subcaption}  
\usepackage{caption}     
\usepackage{gensymb}
\usepackage{amsmath}
\usepackage{amssymb}
\usepackage[colorlinks=true, allcolors=blue]{hyperref}
\journalname{Space Weather}

\begin{document}

%
%


\title{MP-DualNet: A Dual-Branch Deep Learning Approach for Magnetopause Boundary Detection Using Simulated SXI Images}

%
%




\authors{
Cheng Zhong\textsuperscript{+}\affil{1,2},%
Zhongwei Yang\textsuperscript{+}\affil{2},%
Tianran Sun\affil{2},
Chengyan Huang\affil{1},
Jiajia Liu\affil{1,3},
Jingnan Guo\affil{1,3},
Yuming Wang\affil{1,3},
Chi Wang\affil{2}
}


\affiliation{1}{National Key Laboratory of Deep Space Exploration/School of Earth and Space Sciences, University of Science and Technology of China, Hefei 230026, China}

\affiliation{2}{State Key Laboratory of Solar Activity and Space Weather, National Space Science Center, Chinese Academy of Sciences, Beijing 100190, China}

\affiliation{3}{CAS Center for Excellence in Comparative Planetology/CAS Key Laboratory of Geospace Environment/Mengcheng National Geophysical Observatory, University of Science and Technology of China, Hefei 230026, China}






\correspondingauthor{Jiajia Liu}{jiajialiu@ustc.edu.cn}




\begin{keypoints}
\item A new method is developed to automatically detect the Earth’s magnetopause in soft X-ray images with limited field-of-view coverage.
\item The method identifies the magnetopause boundary whether it is fully visible within or partially outside the SXI field of view.
\item The method enables robust magnetopause tracking under varying noise, supporting future SMILE space weather studies.
\end{keypoints}

%
%

%
%


\begin{abstract}
The Earth’s magnetopause is a key boundary governing solar wind–magnetosphere coupling. 
The Solar wind Magnetosphere Ionosphere Link Explorer (SMILE) mission will provide global soft X-ray imaging of the magnetosphere through solar wind charge exchange (SWCX) emission, offering a new opportunity for remote sensing of magnetopause dynamics. However, the limited field of view (FOV) and variable observation geometry of the Soft X-ray Imager (SXI) pose challenges for reliable magnetopause detection. In this study, we develop MP-DualNet, an end-to-end dual-branch deep learning framework for automated magnetopause detection from simulated SXI observations. The framework jointly performs coarse localization to determine magnetopause visibility and relative position within the FOV, and fine-scale boundary localization through pixel-level segmentation. The model is trained and evaluated using line-of-sight magnetopause envelope maps derived from global hybrid simulations under different solar wind and interplanetary magnetic field conditions.
The results demonstrate stable and consistent performance, with a classification accuracy exceeding 0.94 and a segmentation IoU above 0.97. For the test set, mean absolute angular displacement from the reference boundary was $0.488^\circ$. The model maintains robust performance across a wide range of noise levels and generalizes well, as demonstrated by its application to an independent storm event. These results indicate that MP-DualNet provides a reliable and generalizable approach for magnetopause detection in SXI observations, supporting future 3D reconstruction and space weather studies in the context of the SMILE mission.
\end{abstract}

\section*{Plain Language Summary}

Earth's magnetic field creates a protective region around the planet. Its outer boundary, the magnetopause, moves inward and outward as space weather conditions in the solar wind change. The Solar wind Magnetosphere Ionosphere Link Explorer (SMILE) spacecraft will observe this boundary through the  soft X-ray produced near Earth. Interpreting these images is difficult because the boundary may be only partly visible or outside the image, and weak signals can be obscured by  noise. In this work, we developed an automated method that first determines whether the magnetopause is visible and then traces its location when it appears in the image. The method was trained and tested with synthetic images generated from simulations of the solar wind interacting with Earth's magnetic environment. In tests, it correctly identified the boundary's position relative to the image in about 95\% of cases. Among images where the boundary should have been visible, the average difference for those traces was roughly half a degree. The method also captured the overall boundary motion during a separate simulated magnetic storm. These results will support the automated analysis of future SMILE images, although performance must be reassessed with real observations after they become available.

\section{Introduction} \label{sect:intro}

Earth’s magnetopause (MP) is the primary boundary formed by the interaction between the solar wind and Earth’s intrinsic magnetic field. It separates the shocked solar wind plasma in the magnetosheath from the plasma confined within the magnetosphere \citep{geomagmp1963sonettabrams,geomagneticboundary1963cahillamazeen,mp_detection_dl_lownsw_2023zhang,mploc_statisticalinv_sxi_2024cucho,mp3dxray_emission_cmem_jgr2024wharton,mpboundary_dip_denoisemodel_sxi_2025wei}. As a key dynamical interface in the Sun–Earth system, the magnetopause position and global morphology vary with upstream solar wind conditions, especially the solar wind dynamic pressure and the orientation and strength of the interplanetary magnetic field (IMF) \citep{mploc_statisticalinv_sxi_2024cucho, 2024storm_mpcusps_sxi_2025ng}. These spatiotemporal changes directly reflect the evolving solar wind–magnetosphere coupling and the resulting large-scale magnetospheric response \citep{shue1998magnetopause,staples2020magnetopause}. During extreme space weather events, such as coronal mass ejections (CMEs), the magnetopause can undergo rapid earthward motion and substantial compression {\citep{shue1998magnetopause,liu2014qhcmes}. As the primary interface governing solar wind energy and mass transfer into the magnetosphere, its dynamic response under storm conditions provides critical insight into solar wind–magnetosphere coupling processes\citep{Buzulukova2022StromEventReview}.

Since the magnetopause was first identified by spacecraft observations in the 1960s, in-situ missions (e.g., MMS, Cluster, GOES, and THEMIS) have provided major insights into the magnetopause microstructure, boundary layer processes, and transient dynamics \citep[e.g.,][]{globalsi_sxi_deformations_2023yang,guo2024magnetopausedeformation,smile_xrayuv_modeling_2025_review}. Nevertheless, in-situ measurements are intrinsically constrained by single-point sampling or sparsely distributed multi-spacecraft constellations, which limit their ability to continuously resolve the global geometry and evolution of the magnetopause. As a result, in-situ observations alone cannot fully capture the global-scale magnetospheric response across a broad range of solar wind conditions \citep{mploc_statisticalinv_sxi_2024cucho}. Solar wind charge exchange (SWCX) provides a physical mechanism for remote sensing of the terrestrial magnetosphere. In SWCX reactions, highly charged heavy ions in the solar wind (e.g., O$^{6+}$, O$^{7+}$, O$^{8+}$, Fe$^{18+}$, and Mg$^{12+}$) interact with exospheric neutral species (primarily H) \citep{Zhang2022SWCX_ICME}. An electron is captured into an excited state of the ion, followed by radiative cascades that produce characteristic line emission from the extreme ultraviolet (EUV) to the soft X-ray bands \citep{cometxraysource_sw_1997cravens,solarsystemxrays2007bhardwaj,4methos_bfa_tfa_tda_cta_gl2022wang}. 

SWCX emission in the near-Earth environment is typically enhanced in the magnetosheath, where the solar wind ion flux is high and the line-of-sight (LOS) neutral column density, mainly supplied by the geocorona, is sufficient to produce strong soft X-ray emission. The resulting emissivity pattern is statistically correlated with large-scale boundaries, including the bow shock and the magnetopause \citep{swcx2003robertson,magnetospherexray2006robertson,Sun2019_MHD_Magnetopause,mpboundary_dip_denoisemodel_sxi_2025wei,conditions_pressure_imf_2025varela}. Soft X-ray imaging based on SWCX is therefore regarded as a direct means of diagnosing the global configuration and temporal evolution of the magnetosphere, complementing traditional in-situ observation for studies of global-scale solar wind-magnetosphere coupling \citep{bfa2019,mp3dxray_emission_cmem_jgr2024wharton}.

SMILE is a joint mission between the Chinese Academy of Sciences (CAS) and the European Space Agency (ESA). The spacecraft was successfully launched aboard a Vega-C rocket on 19 May 2026 and subsequently entered its planned highly elliptical Earth orbit. It will operate in a highly elliptical polar orbit with a perigee of $\sim 2~R_E$, an apogee of $\sim 19.8~R_E$, and an orbital period of $\sim 50.5$~h. Each orbit provides an extended viewing interval of the dayside magnetosheath and cusp regions. The nominal continuous soft X-ray observing time is about 41~h per orbit \citep{sxi_mission_short_2024_review,sxi_mission_long_2025_review,smile_xrayuv_modeling_2025_review,SMILEssrv2025WangChi}. 
One of the primary scientific objectives of the SMILE mission is to investigate the large-scale structure of Earth’s magnetosphere using soft X-ray imaging and to characterize the global modes of solar wind–magnetosphere interactions \citep{Carter2025SMILEEngagement}. It will carry a Soft X-ray Imager (SXI), which employs lobster-eye optics to obtain the Soft X-ray images of the interaction between the solar wind and the magnetosphere. SXI has a large field of view (FOV) of approximately $16^{\circ}\times 26^{\circ}$. Within this framework, accurately determining the magnetopause location and geometry from SXI images is a fundamental requirement to help accomplish the scientific goals of SMILE. Because reliable extraction of the magnetopause boundary provides the basis for three-dimensional (3D) reconstruction of the magnetosphere and for quantitative analyses of its dynamical response to solar wind \citep{mp3dxray_emission_cmem_jgr2024wharton}.

At present, methods for magnetopause detection based on soft X-ray imaging can be broadly classified into two categories: traditional approaches and machine learning approaches. Traditional methods comprise several representative schemes, each of which relies on specific physical assumptions and observational conditions, such as prescribed magnetopause geometries, viewing geometries, or simplified emission models \citep{4methos_bfa_tfa_tda_cta_gl2022wang,mploc_statisticalinv_sxi_2024cucho}. Five of the most commonly used traditional methods are briefly listed below. Readers are encouraged to find their details in the corresponding references.

The first method is the Tangential Direction Approach (TDA) \citep{tda2018}. This method is based on the assumption that the peak soft X-ray emission corresponds to the tangential direction of the magnetopause \citep{mp_detection_dl_lownsw_2023zhang}. By using two SXI images acquired under identical upstream solar wind conditions, the tangential directions and corresponding tangency points of the magnetopause are inferred. However, TDA requires quasi-stationary solar wind conditions during the observation interval, which limits its applicability under dynamically varying solar wind environments. 
The second method is the Boundary Fitting Approach (BFA) \citep{bfa2019}. In this method, parameterized geometric models of the magnetopause and bow shock are constructed and fitted to the observed SXI images through a global optimization procedure. 
The optimal model parameters are obtained by minimizing a cost function defined by the discrepancy between the modeled and observed soft X-ray emission intensities. Due to its strong dependence on the intensity distribution, the BFA method is sensitive to noise contamination and exhibits significantly degraded performance under low signal-to-noise ratio (SNR) conditions. 
The third method is the Tangential Fitting Approach (TFA) \citep{tfa2020}, which combines key features of both BFA and TDA. TFA reconstructs the magnetopause geometry by matching the tangential directions extracted from SXI images with those derived from a parameterized magnetopause model. Compared with BFA, TFA involves a reduced number of free parameters, typically three, thereby lowering the model complexity. Nevertheless, this approach still relies on prior assumptions regarding the functional form of the magnetopause shape. 
The fourth method is the Computerized Tomography Approach (CTA) \citep{cta2022}. 
This method reconstructs the 3D magnetopause structure using multiple 2D SXI images acquired from different viewing perspectives. An important advantage of CTA is that it does not require a predefined parameterization of the magnetopause geometry. However, CTA assumes that the upstream solar wind conditions remain approximately constant over the multiple observation epochs, a requirement that is difficult to satisfy in realistic space weather conditions. 
A statistical inversion approach was subsequently developed by \citet{mploc_statisticalinv_sxi_2024cucho}. Their method applies maximum a posteriori (MAP) estimation to soft X-ray LOS measurements, together with the viewing geometry and a prior emissivity model, to reconstruct the 3D magnetosheath emissivity distribution. The magnetopause location along the Sun--Earth line is then inferred from the reconstructed emissivity. Unlike TFA, which fits tangent directions with a parameterized magnetopause model, this approach estimates the emitting plasma distribution before deriving the boundary location. Its reconstruction accuracy depends on the observing geometry and the prior emissivity model.

Overall, the BFA and TFA methods are capable of reconstructing the magnetopause using a single SXI image, but they rely on strong prior assumptions regarding the magnetopause geometry \citep{4methos_bfa_tfa_tda_cta_gl2022wang}. 
In contrast, the TDA and CTA methods require multiple SXI observations and assume quasi-stationary upstream solar wind conditions over the observation interval, which restricts their applicability during dynamic solar wind variations. 
Statistical inversion provides 3D emissivity and standoff-distance information but requires an explicit viewing geometry and prior emissivity model.
It is worth noting that magnetic storms are a primary scientific focus of the SMILE mission. During these dramatic events, the magnetopause undergoes rapid and large-scale displacement, frequently moving outside the SXI FOV. Traditional methods, which predominantly rely on the entire boundary being visible within the FOV, lack a systematic and robust solution for such incomplete signature scenarios.
To improve the applicability of conventional reconstruction methods, machine-learning techniques have been introduced to assist reconstruction. For example, \citet{fov_cls_mp_rf_2023wang} developed the UCFE-RF framework, which combines unsupervised feature extraction with a random forest classifier to assess whether simulated magnetospheric soft X-ray images contain sufficiently clear and complete magnetopause signatures for subsequent reconstruction. The machine-learning component therefore performs image-level classification and pre-screening rather than directly determining the magnetopause boundary or reconstructing its 3D structure. When the relevant boundary signatures are adequately captured within the SXI FOV, the selected images can subsequently be processed using the TFA method to reconstruct the 3D dayside magnetopause.

Unlike machine-learning-assisted approaches, direct localization methods aim to infer the magnetopause position and boundary from the SXI image itself. From an image processing perspective, magnetopause localization in SXI images can be naturally formulated as a semantic segmentation problem, in which the spatial and intensity characteristics of soft X-ray emissions are learned to distinguish pixel regions located inside and outside the magnetopause. With the rapid development of deep learning techniques, semantic segmentation methods based on convolutional neural network (CNN) have demonstrated strong representation capability in high-dimensional feature space, providing an effective way for automated identification of magnetopause boundaries \citep{cnn2018,deeplab2018,fcn2018,pan2019,resunetpp2019,pspnet2021}. In recent years, several studies have explored the application of deep learning to magnetopause detection using SXI observations. \citet{mp_detection_dl_lownsw_2023zhang} focused on low solar wind density conditions and trained a neural network to identify photon count peak locations in SXI images. The pixel-level localization accuracy reached 90.94\%, and the subsequent application of the TFA method yielded a subsolar magnetopause position error smaller than 0.2~$R_E$. \citet{oesa-unet_att_limitedfov_2024wang} trained an attention-enhanced OESA-UNet model using simulated soft X-ray images and corresponding magnetopause masks. The model achieved a precision of 92.1\%. Its average angular deviation from the reference boundary was reported below $0.005^\circ$.

Nevertheless, in a realistic space observation environment, background noise and weak soft X-ray emission are ubiquitous, posing significant challenges to robust extraction and stable identification of magnetopause boundaries. 
To address this issue, \citet{mpboundary_dip_denoisemodel_sxi_2025wei} systematically investigated cases with blurred magnetopause signatures under low SNR conditions and proposed an optimization framework that combines image calibration with a denoising model based deep image prior. Under moderate to high solar wind density conditions ($N = 10$--$20~\mathrm{cm^{-3}}$), the normalized mean squared error (NMSE) of the reconstructed magnetopause boundary can be controlled below $0.05$.

In this study, we propose an end-to-end dual-branch deep learning framework, termed MP-DualNet, for automated detection and precise localization of the magnetopause from simulated SXI images under a range of different SNR conditions. The proposed framework is designed to determine whether the magnetopause enters the SXI FOV, to distinguish its relative orientation within the FOV, and to further extract the pixel-level location of the magnetopause boundary. 

MP-DualNet adopts a dual-branch architecture with two separate encoders. The backbone network is based on a U-Net architecture, in which low-level spatial details and high-level semantic information are integrated through skip connections between the downsampling and upsampling paths. One branch serves as a classification model for coarse localization, predicting the relative position of the magnetopause with respect to the SXI FOV. The other branch acts as a segmentation model, producing a pixel-wise probability map of the magnetopause region. By combining the outputs of the classification and segmentation models and applying a binary threshold, a final mask representation of the magnetopause boundary is obtained.

The remainder of this paper is organized as follows. 
Section~\ref{sec_data} describes the generation of simulated SXI images, the construction of the dataset, and the overall architecture of MP-DualNet. 
Section~\ref{sec_method} presents the training configuration, including hyperparameter settings and evaluation metrics. 
Section~\ref{sec_expsAndmetrics} reports the experimental results and provides assessments of the model performance in both classification and segmentation tasks. Section~\ref{sec_results} discusses the robustness and generalization capability of the proposed method under different noise levels and observational conditions. 
Section~\ref{sec_discussion_conclusion} presents the discussions and conclusions.

\section{Data} \label{sec_data}
The dataset used in this study was constructed before the launch of SMILE and consists entirely of simulated SXI observations \citep{sxi_mission_long_2025_review}. The overall dataset construction procedure is summarized as follows. First, global magnetospheric numerical simulations are employed to model the SWCX process, yielding 3D distributions of soft X-ray volumetric emissivity under different upstream solar wind conditions, spacecraft orbital configurations, and observation geometries. Then, the simulated emissivity distributions are used as inputs to an SXI forward modeling framework, in which LOS integration is performed to generate 2D SXI images corresponding to specific viewing directions \citep{mp_detection_dl_lownsw_2023zhang}.

\subsection{SXI Simulation} \label{subsec_sxi_sim}
Three-dimensional global hybrid simulations are widely used to describe the large-scale structure and dynamical evolution of the terrestrial magnetosphere under varying solar wind conditions \citep{hybridcodes2023winske,globalhysimswandms2014lin,palmroth2023magnetotaileruptions,fatemi2024magnetosheathjets,grandin2024hybridvlasovxray,globalsi_sxi_deformations_2023yang}. Variations in solar wind dynamic pressure and IMF conditions lead to prominent changes in the location and morphology of key magnetospheric structures, including the bow shock, magnetopause, and cusp regions \citep{Yang2016GlobalPIC,mixingrete_mpboundary_khvortices,2024storm_mpcusps_sxi_2025ng}.

In this study, SWCX emission associated with solar wind-magnetosphere interaction is modeled based on global hybrid simulation results, yielding the 3D soft X-ray volumetric emissivity distribution $P_x$ \citep{smile_xrayuv_modeling_2025_review}. For a given spacecraft orbital position and viewing geometry, the volumetric emissivity is integrated along the SXI instrument's LOS to obtain the 2D LOS soft X-ray intensity distribution. The corresponding expression is given by \citet{mploc_statisticalinv_sxi_2024cucho}.
\begin{equation}
I(\mathbf{r}, \hat{\mathbf{n}}, t)
=
\frac{1}{4\pi}
\int_{0}^{L_{\max}}
P_x(\mathbf{r}, t)\,\mathrm{d}l
,
\label{SXI_Intensity}
\end{equation}
where $I(\mathbf{r}, \hat{\mathbf{n}}, t)$ denotes the time-dependent soft X-ray energy flux (in units of keV\,cm$^{-2}$\,s$^{-1}$\,sr$^{-1}$) observed along the viewing direction $\hat{\mathbf{n}}$ at geocentric position $\mathbf{r}$ and time $\mathrm{t}$. Here, $l$ denotes the distance along the LOS, $L_{\max}$ is the maximum integration length.

The simulated SXI images cover a FOV of approximately $15.5^{\circ} \times 25.5^{\circ}$, with an angular resolution of $1^{\circ}$ in both the horizontal and vertical directions, corresponding to a total of 375 pixels. The soft X-ray intensity maps derived from the hybrid simulations are constructed on the same angular grid. 
Each synthetic SXI image was generated from a single simulation snapshot. The integration in Equation~(\ref{SXI_Intensity}) is performed along the line of sight and does not represent temporal
averaging over an instrumental exposure. Accordingly, no detector integration time was assigned to the images used in this study.

In this work, three representative sets of upstream solar wind and IMF conditions are selected to span a range of typical and disturbed solar wind states. These three cases are designed to sample different combinations of solar wind velocity, density, temperature, and IMF orientation, which are known to strongly influence the global magnetospheric configuration and the location of the magnetopause \citep{conditions_pressure_imf_2025varela}. Specifically, the selected runs include a high-speed, moderate-density case with northward IMF (Run 1), a high-density case with strongly southward IMF (Run 2), and a lower-speed, low-density case with weakly inclined IMF (Run 3). Together, these scenarios allow us to assess the performance of the proposed method under diverse magnetospheric conditions, including both relatively steady and strongly disturbed states. For each run, the 3D soft X-ray volumetric emissivity distribution is obtained from global hybrid simulations. The corresponding LOS soft X-ray intensity distributions are then computed using Equation~(\ref{SXI_Intensity}) for specified spacecraft orbital positions and viewing geometries. The observer positions and pointing directions were selected from the nominal SMILE orbit. The upstream solar wind and IMF parameters adopted in these three global magnetospheric simulations are summarized in Table~\ref{tab:sw_imf_params}. Specifically, the selection of solar wind densities in Runs 1 and 2 ($N_\mathrm{sw} = 20$ and $60 \text{ cm}^{-3}$, respectively) aligns with the SMILE instrument team’s recommendation that $N_\mathrm{sw} \ge 15 \text{ cm}^{-3}$ is optimal for soft X-ray imaging. These high-density scenarios mimic CME-sheath conditions, providing sufficient SWCX emission for clear boundary identification and serving as a critical pre-study for storm-time events. In contrast, Run 3 ($N_\mathrm{sw} = 7 \text{ cm}^{-3}$) represents a lower-density environment where blurred signatures pose a greater challenge for automated detection, thereby testing the model's robustness under less-than-ideal observational conditions.

\begin{table}[htpb]
\caption{Upstream solar wind and interplanetary magnetic field (IMF) parameters adopted in the global magnetospheric simulations.}
\label{tab:sw_imf_params}
\centering
\begin{tabular}{cccccc}
\hline
Run & $V_{\mathrm{sw}}$ (km\,s$^{-1}$) & $N_{\mathrm{sw}}$ (cm$^{-3}$) & $T_{\mathrm{sw}}$ (K) & IMF $\mathbf{B}$ (nT, GSM) \\
\hline
1 & $-800$ & 20 & $1.5\times10^{5}$ & $(0, 0, 6)$ \\
2 & $-750$ & 60 & $5.0\times10^{5}$ & $(0, -40, -40)$ \\
3 & $-450$ & 7  & $1.5\times10^{5}$ & $(-9.96, 0.6, 0.6)$ \\
\hline
\end{tabular}
\begin{flushleft}
\footnotesize
\textbf{Note.} $V_{\mathrm{sw}}$ denotes the upstream solar wind velocity, $N_{\mathrm{sw}}$ the solar wind number density, and $T_{\mathrm{sw}}$ the solar wind temperature. 
The IMF vector $\mathbf{B}$ is given in GSM coordinates.
\end{flushleft}
\end{table}

\subsection{Noise Modeling} \label{subsec_noiseModeling}

The noise-free images used in this study are LOS-integrated soft X-ray intensity maps rather than detector photon-count maps. Accordingly, the added noise was not generated using a Poisson process. Instead, a controlled additive random background was introduced to evaluate the sensitivity of MP-DualNet to different levels of image perturbation.

For a noise-free intensity image \(I^{(0)}\), the image at noise-amplitude level \(A_k\) and random realization \(r\) was generated as
\begin{equation}
I_{ij}^{(k,r)}
=
I_{ij}^{(0)}
+
A_k U_{ij}^{(k,r)},
\label{eq:additive_noise}
\end{equation}
where \(i\) and \(j\) denote the image pixels and \(U_{ij}^{(k,r)}\) was generated uniform distribution. Therefore, $U_{ij}^{(k,r)}\sim\mathcal{U}(0,1).$ The resulting additive term is uniformly distributed over \([0,A_k]\), with mean \(A_k/2\) and variance \(A_k^2/12\). Random values were generated independently for different pixels and realizations.

Ten amplitude parameters were considered:
\begin{equation}
A_k \in
\{5,10,15,20,25,30,35,40,45,50\}
~\mathrm{keV\,cm^{-2}\,s^{-1}\,sr^{-1}}.
\end{equation}
For each amplitude, ten independent random realizations were added to the same noise-free image. Together with the original noise-free image,
this procedure produced 101 images for each source-image configuration. All images derived from the same noise-free source image retained the
same reference magnetopause label.

This additive-noise model provides a controlled test of robustness to increasing random background contamination. It is not intended to
reproduce the complete noise statistics of SXI observations. In particular, it does not include Poisson photon-counting statistics, the
instrument response, or a physical model of the spatial and spectral properties of the cosmic X-ray background. Detector-level evaluation
will require the use of the SXI response and exposure settings to calculate expected photon counts, followed by Poisson sampling and
validation against flight observations.

\subsection{Target Annotation}
Based on the result of the global hybrid simulations, the reference ground truth for the magnetopause is constructed in the form of a LOS magnetopause envelope, which represents the projected envelope of the magnetopause boundary along the viewing direction \citep{yang2025b}. In this study, the mixing rate (MR) is adopted as the indicator variable, whose physical meaning and formulation are given below. The grid cells corresponding to the magnetopause boundary are then projected along the LOS under specified spacecraft positions and viewing geometries, and an envelope extraction procedure is applied to obtain a 2D LOS magnetopause envelope. The resulting LOS magnetopause envelope is consistent with the simulated SXI images in both spatial extent and angular resolution. The mixing rate is defined as Equation~(\ref{mixing_rate}).
\begin{equation}
\mathrm{MR}
=
-2 \left| \frac{N_{\mathrm{sw}}}{N_{\mathrm{tot}}} - 0.5 \right| + 1,
\label{mixing_rate}
\end{equation}
where $N_{\mathrm{sw}}$ and $N_{\mathrm{ms}}$ denote the number densities of solar wind plasma and magnetospheric plasma, respectively, and $N_{\mathrm{tot}} = N_{\mathrm{sw}} + N_{\mathrm{ms}}$. This definition is mathematically equivalent to the alternative mixing fraction formulation used in previous studies. An important advantage of the mixing rate is that it is bounded within the interval $[0,1]$, producing a pronounced peak at the boundary layer where plasma mixing is strongest \citep{Matsumoto2006KHTurbulence,daughton2014reconnectionrate}. Accordingly, this quantity provides a robust and automated criterion for identifying the magnetopause boundary in the global hybrid simulations and is adopted throughout this work.

Within the SXI FOV, the LOS magnetopause envelope divides the observation region into two topologically connected domains, corresponding to the solar wind–magnetosheath side and the magnetospheric side. This partition is binarized to form pixel-level segmentation labels, in which pixels located on the magnetospheric side are assigned a value of 1, and those on the solar wind side are assigned a value of 0. Under this formulation, magnetopause boundary identification is cast as a semantic segmentation task, aiming to recover the pixel-wise mask of the magnetopause from SXI images.

Considering that the magnetopause does not always fully reside within the SXI FOV \citep{oesa-unet_att_limitedfov_2024wang}, an additional coarse localization label is assigned to each sample to reflect the relative position of the magnetopause envelope with respect to the FOV. 
Three discrete categories are defined: Left, Inside, and Right. The labels Left and Right indicate cases in which the magnetopause envelope lies predominantly outside the left or right boundary of the FOV, respectively, whereas Inside corresponds to cases in which the magnetopause envelope is fully contained within the FOV. These coarse localization labels are used to support the design of the classification model in the proposed network, enabling stable magnetopause localization when the boundary is only partially visible or lies outside the FOV.

In this study,  noise is superimposed on noise-free soft X-ray intensity images solely to modify the SNR. The underlying magnetopause geometry, which is prescribed by the hybrid simulations, remains unchanged. As a result, for a given noise-free soft X-ray image, all SXI realizations generated at different noise levels share an identical LOS magnetopause envelope segmentation label. This design preserves the physical consistency of the ground truth and allows for a controlled and unbiased evaluation of model robustness across varying noise conditions.

\section{Methods} \label{sec_method}

\subsection{Overview of the MP-DualNet Framework}

MP-DualNet is an end-to-end dual-branch deep learning framework designed for robust detection and precise localization of the magnetopause from simulated SXI images. The framework takes SXI images as input and jointly addresses two related but distinct tasks: magnetopause visibility identification and boundary localization. This formulation enables explicit task decoupling while maintaining consistency in feature representation.

\begin{figure}[htbp]
    \centering
    \includegraphics[width=0.6\linewidth]{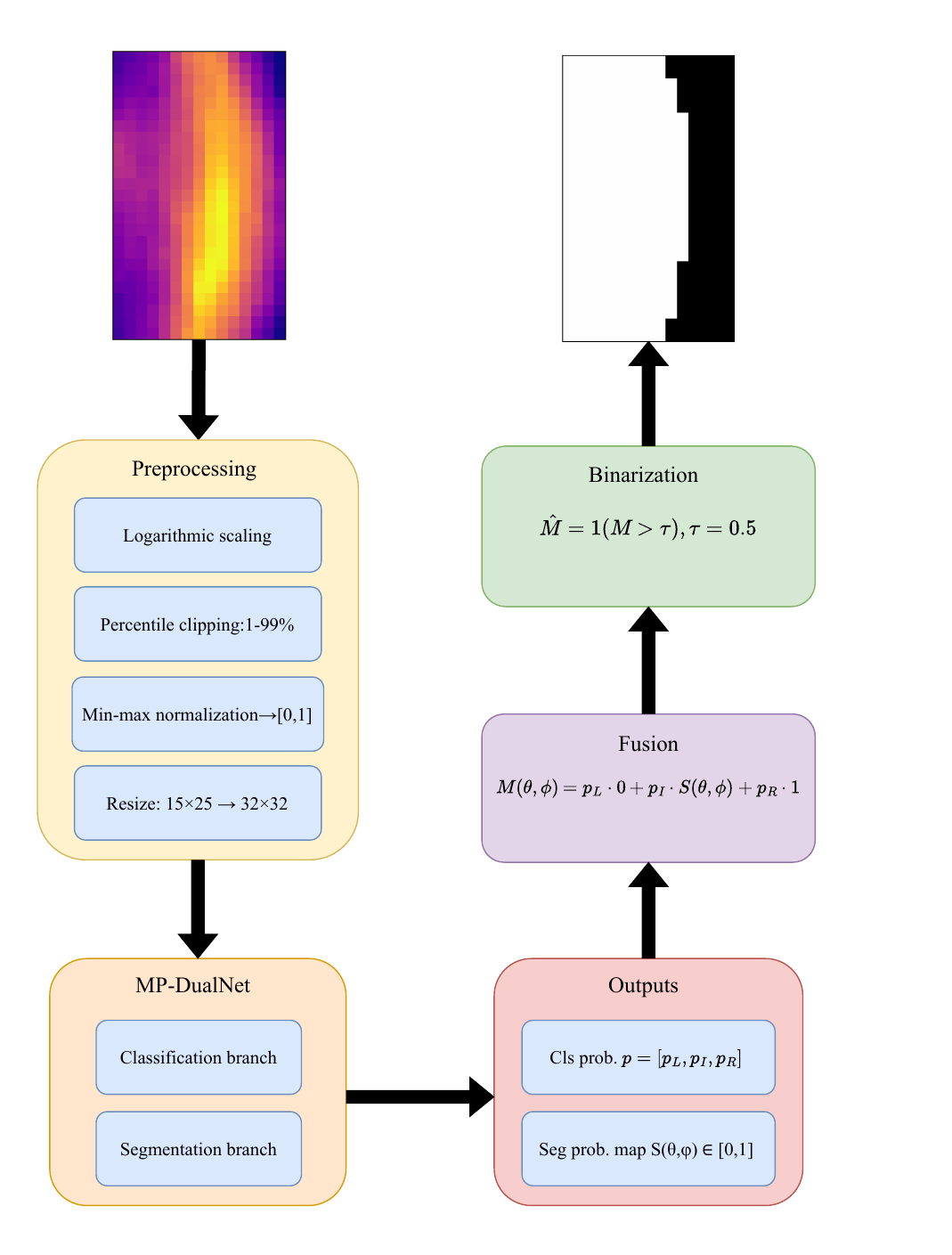}
    \caption{Overall flowchart of magnetopause detection using the MP-DualNet framework. A detailed explanation of the framework's structure is provided in the main text.}
    \label{fig:flowchart}
\end{figure}

\begin{figure}[htbp]
    \centering
    \includegraphics[width=1.0\linewidth]{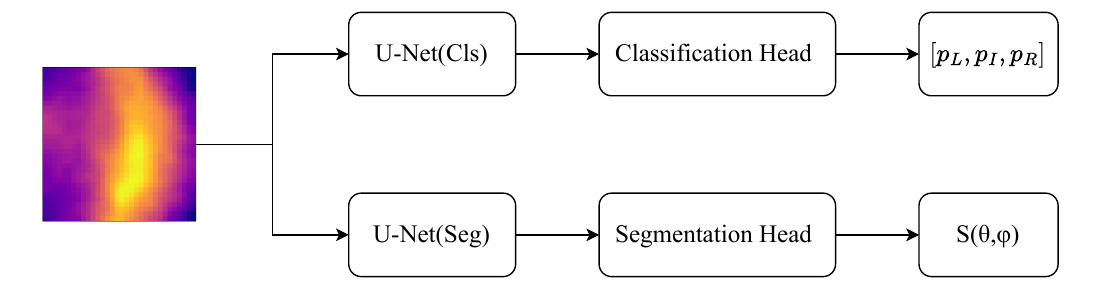}
    \caption{Network architecture of MP-DualNet. The framework consists of parallel classification and segmentation branches. U-Net (Cls) and U-Net (Seg) serve as the respective backbones for the classification and segmentation tasks. $[p_L, p_I, p_R]$ denotes the probabilities for the magnetopause location being to the left, inside, or to the right of the SXI field of view. $\mathrm{S}(\theta, \phi)$ represents the segmentation map in angular coordinates.}
    \label{fig:mpdualnet_arch}
\end{figure}

As illustrated in Figure~\ref{fig:flowchart}, the input SXI images are first subjected to preprocessing operations tailored to the high dynamic range and noise characteristics of soft X-ray observations. These steps enhance intensity gradients associated with the magnetopause and other large-scale magnetospheric structures. The preprocessed images are then fed into a U-Net shape feature extraction network, which captures multiscale spatial information and high-level semantic features through a hierarchical encoder–decoder architecture with skip connections \citep{resunetpp2019,oesa-unet_att_limitedfov_2024wang,unetjet2024liujj}. Based on the extracted feature representations, MP-DualNet adopts a dual-branch design to perform coarse and fine localization tasks (Figure~\ref{fig:mpdualnet_arch}). The classification branch is responsible for determining the magnetopause's visibility within the SXI FOV and its relative orientation. The segmentation branch produces a pixel-wise probability map of the magnetopause envelope when the boundary is observable within the FOV. 

By explicitly decoupling magnetopause visibility classification from boundary segmentation, MP-DualNet remains effective in complex observational scenarios, including cases in which the magnetopause is partially or entirely outside the FOV or when boundary signatures are weak under low SNR conditions. The final magnetopause envelope mask is generated by combining the outputs of the classification and segmentation branches. 
This output provides a reliable input for subsequent 3D magnetopause reconstruction and physics-based analysis using soft X-ray observations.

\subsection{Training Dataset and Preprocessing}
As described in Section~\ref{subsec_sxi_sim} and summarized in Table~\ref{tab:sw_imf_params}, three representative sets of upstream solar wind and IMF conditions are used to generate the global magnetospheric simulations. These different upstream conditions lead to distinct magnetopause configurations, resulting in systematic variations in the relative position of the magnetopause with respect to the FOV of SXI.

Based on the simulated SXI observations derived from these three runs, the dataset is categorized into three observation types according to the relative location of the magnetopause envelope in the FOV of SXI. Specifically, the magnetopause envelope may be located predominantly on the left side of the SXI FOV (Left), fully contained within the FOV (Inside), or located predominantly on the right side of the FOV (Right).

To mitigate potential bias arising from class imbalance, a balanced sampling strategy is adopted, with each category containing 4,747 samples, as summarized in Table~\ref{tab:dataset_split}. The complete dataset is split into training, validation, and test sets at an 8:1:1 ratio. The training set is used for model optimization, the validation set for hyperparameter tuning and overfitting control, and the test set exclusively for final performance evaluation. To prevent data leakage, the noisy and noise-free counterparts are strictly assigned to the same subset. This ensures a more robust and unbiased assessment of generalization capability. The three classes were sampled equally to prevent the training objective from being dominated by one FOV state and to evaluate each class with equal statistical weight. Each class contains 3,737 training samples, 505 validation samples, and 505 test samples. This balanced distribution is a model-development choice and is not intended to represent the frequencies of the three classes during SMILE operations.

\begin{table}[t]
\centering
\caption{Number of samples in each subset and class used for dataset splitting.}
\label{tab:dataset_split}
\begin{tabular}{lcccc}
\hline
Split & Left & Right & Inside & Total \\
\hline
Training   & 3737 & 3737 & 3737 & 11211 \\
Validation &  505 &  505 &  505 &  1515 \\
Test    &  505 &  505 &  505 &  1515 \\
\hline
\end{tabular}
\end{table}
Neural networks possess strong function approximation capability and are theoretically capable of learning arbitrary nonlinear mappings \citep{MultilayerNetworks1989Hornik}. In conjunction with the noise modeling strategy described in Section~\ref{subsec_noiseModeling}, both noise-free and noise-contaminated SXI images are included during training and are assigned identical magnetopause envelope labels. 
For each noise-free SXI image, multiple noise realizations are generated and treated as distinct samples. This design explicitly encourages the network to focus on the structural characteristics of the magnetopause rather than adapting to specific noise patterns or intensity distributions, and it enables a fair and robust assessment of model generalization under realistic observational conditions.


Given the large dynamic range and spatially nonuniform noise characteristics of SXI images, a global min-max normalization across the entire dataset is not adopted. Instead, an adaptive preprocessing strategy based on individual images is employed. For each SXI image, the intensity is first transformed using logarithmic scaling. The 1 and 99 percentiles, denoted as $P_{1}$ and $P_{99}$, are then computed from the image itself, and pixel values are clipped to the interval $[P_{1}, P_{99}]$ to suppress extreme outliers and saturated pixels. The clipped image is subsequently linearly rescaled to the range $[0,1]$, yielding the normalized intensity.
\begin{equation}
I_{\mathrm{norm}}
=
\frac{\mathrm{clip}(I, P_1, P_{99})  - P_{1}}{P_{99} - P_{1}},
.
\label{equ_Inorm}
\end{equation}

Previous studies have shown that the magnetopause in SXI images is typically located between the soft X-ray intensity peak and regions of maximum intensity gradient \citep{2method_ixandgrad_mploc_2022samsonov}. The adopted preprocessing strategy enhances image contrast while preserving gradient information, enabling the model to identify magnetopause boundaries based on structural and gradient-related features rather than absolute intensity levels across different solar wind and IMF conditions.

\subsection{Backbone Network}
U-Net is a CNN widely used for image segmentation tasks \citep{unet2015ronneberger}. The backbone network adopted in this study is based on a U-Net architecture, which serves as the core feature extractor for both magnetopause localization and boundary segmentation. The U-Net architecture is particularly suitable for SXI image analysis because it combines multi-scale contextual information with fine spatial detail through an encoder--decoder structure with skip connections.

Figure~\ref{fig:unet} shows the detailed architecture of the U-Net backbone used in this work. The network takes a single-channel SXI image as input and consists of a contracting path (encoder) and an expanding path (decoder). In the encoder, feature extraction is performed across multiple spatial scales through successive convolutional blocks followed by downsampling. Each convolutional block contains two $3\times3$ convolution layers with ReLU activation, and spatial resolution is reduced by a $2\times2$ max-pooling operation. The number of feature channels increases progressively from 16 to 256 as the spatial resolution decreases, enabling the network to capture increasingly abstract and global representations of the magnetopause structure.

At the bottleneck layer, the feature maps reach the lowest spatial resolution ($2\times2$) with 256 channels, providing a compact representation that encodes large-scale geometric information. The decoder then restores spatial resolution through successive $2\times2$ upsampling operations.
At each decoder stage, the upsampled feature maps are concatenated with the corresponding encoder features via skip connections. These skip connections allow fine-scale spatial information lost during downsampling to be reintroduced, which is essential for preserving boundary geometry and accurately delineating the magnetopause envelope.

Dropout layers are applied at selected levels to improve generalization and reduce overfitting, particularly under strong  noise conditions. The final decoder output provides a high-resolution feature representation that is subsequently fed into the task-specific classification and segmentation heads described in the following sections.

\begin{figure}[htbp]
    \centering
    \includegraphics[width=0.95\linewidth]{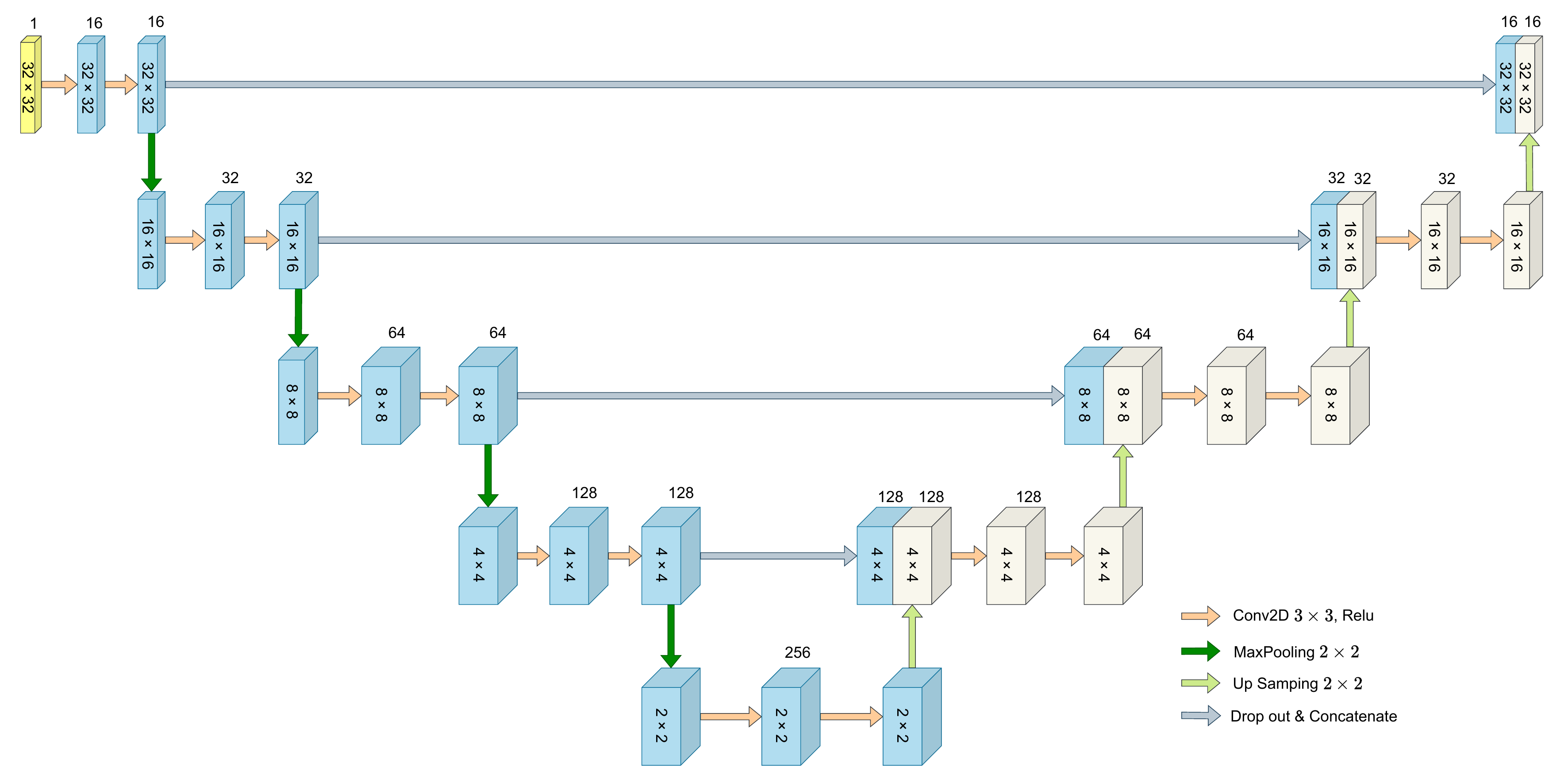}
    \caption{Architecture of the U-Net backbone network used in this study.
The network follows an encoder--decoder structure with five resolution levels. Blue blocks denote convolutional feature maps, with numbers indicating the spatial resolution and channel dimension. Downward green arrows represent $2\times2$ max-pooling operations, upward light-green arrows indicate $2\times2$ upsampling, and horizontal connections denote feature concatenation via skip connections. The overall design is adapted from the U-Net neural architecture of \citet{unetjet2024liujj}, with modifications to channel configuration and regularization to suit simulated SXI images and magnetopause detection.}
\label{fig:unet}
\end{figure}



Although various U-Net variants incorporating attention mechanisms or Transformer-based modules have been proposed \citep{swinunet2022cao}, the relatively simple U-Net architecture is sufficient for the magnetopause detection task considered in this study after several trial-and-error experiments. Given the dataset size and task complexity, this choice reduces model complexity and mitigates the risk of overfitting. More advanced architectures may be explored in future work to further assess potential performance gains.

\subsection{Classification Head for Coarse Localization}

To achieve coarse localization of the magnetopause with respect to the SXI FOV, a classification-specific U-net backbone and a classification head are employed (Figure~\ref{fig:mpdualnet_arch}). As shown in Figure~\ref{fig:cls_head}, the classification branch receives the preprocessed SXI image and extracts the multiscale features associated with the global intensity and distribution and the relative position of the magnetopause with respect to the FOV. Specifically, global average pooling (GAP) is applied to the 32-channel feature maps at the U-Net output, producing a 32-dimensional feature vector. The pooled features are then passed through a fully connected hidden layer (MLP, Multi-Layer Perceptron) with 128 neurons. A softmax activation function is finally applied to output the probabilities $\left[ p_{\mathrm{L}}, p_{\mathrm{I}}, p_{\mathrm{R}} \right]$, corresponding to the cases in which the magnetopause is located on the left side of the FOV, inside the FOV, or on the right side of the FOV, respectively.

\begin{figure}[htbp]
    \centering
    \includegraphics[width=0.95\linewidth]{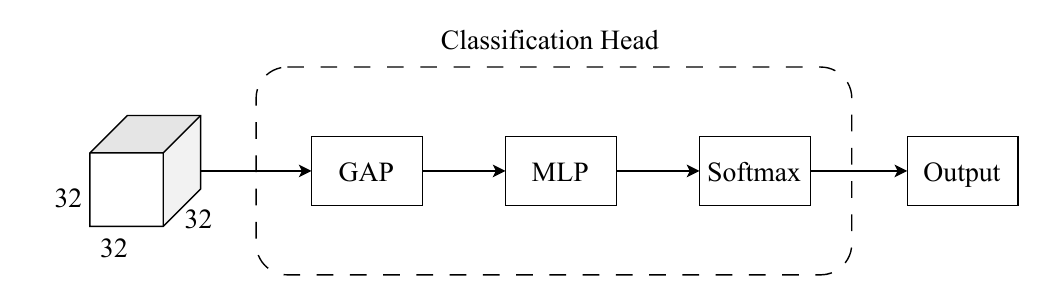}
    \caption{
    Schematic diagram of the classification head used for coarse magnetopause localization. The 32-channel feature maps generated by the classification-specific U-Net backbone are first compressed by global average pooling (GAP) into a 32-dimensional feature vector. The pooled features are then processed by a multilayer perceptron (MLP) for nonlinear feature transformation, followed by a softmax layer that outputs the probability vector 
    $\left[p_{\mathrm{L}}, p_{\mathrm{I}}, p_{\mathrm{R}}\right]$. These probabilities correspond to three relative magnetopause positions with respect to the FOV of SXI.
    }
    \label{fig:cls_head}
\end{figure}

When $p_{\mathrm{I}}$ attains the maximum value, the magnetopause is considered to be observable within the SXI FOV, and the output of the segmentation branch is regarded as valid. Otherwise, the model reports only the relative orientation of the magnetopause.

\subsection{Segmentation Head for Fine Localization}

The segmentation branch is designed for fine localization of the magnetopause and produces a pixel-wise probability map distinguishing the regions inside and outside the magnetopause envelope. In contrast to the image-level classification task, segmentation requires the preservation of the spatial correspondence and fine boundary of information throughout the feature-extraction process. Given the distinct objectives and optimization targets of the classification and segmentation tasks, two independent U-Net backbones are used to serve the respective branches, rather than sharing a common encoder.

The segmentation head consists of a convolutional module applied to the output feature maps of the U-Net backbone. A convolution layer with a kernel size of $7 \times 7$ and a stride of 1 is used to aggregate local spatial context. The convolutional weights are initialized using the Glorot uniform scheme. A sigmoid activation function is applied at the output layer to generate a single-channel probability map representing the likelihood of each pixel belonging to the magnetopause envelope.

\subsection{Dual-branch Magnetopause Detection Architecture}

By integrating the classification and segmentation branches described above, a dual-branch magnetopause detection framework is constructed. The classification branch determines the visibility of the magnetopause within the SXI FOV and predicts its relative orientation. The segmentation branch provides a pixel-wise probability map of the magnetopause envelope when the boundary is observable within the FOV.

The outputs of the two branches are combined to obtain a fused magnetopause-envelope probability map. Specifically, the fused probability map is defined as
\begin{equation}
M(\theta,\phi)
=
p_{\mathrm{L}}\cdot 0
+
p_{\mathrm{I}} S(\theta,\phi)
+
p_{\mathrm{R}}\cdot 1,
\label{eq:fused_probability}
\end{equation}
where \(p_{\mathrm{L}}\), \(p_{\mathrm{I}}\), and \(p_{\mathrm{R}}\) are the classification probabilities for the Left, Inside, and Right classes, respectively, and \(S(\theta,\phi)\) is the raw pixel-wise probability map generated by the segmentation head. The segmentation head always produces \(S(\theta,\phi)\) for every input image. However, its contribution to the fused probability map is controlled by the Inside probability \(p_{\mathrm{I}}\). For images classified as Left or Right, the classification branch provides the dominant coarse FOV information, whereas reliable pixel-level boundary interpretation is mainly associated with Inside cases. When one class probability dominates, the soft fusion naturally approaches the corresponding hard decision }($\hat{c} = \arg\max_{c\in\{\mathrm{L},\mathrm{I},\mathrm{R}\}} p_c$). Specifically, if \(p_{\mathrm{I}}\approx 1\), then \(M(\theta,\phi)\approx S(\theta,\phi)\); if \(p_{\mathrm{R}}\approx 1\), then \(M(\theta,\phi)\approx 1\); and if \(p_{\mathrm{L}}\approx 1\), then \(M(\theta,\phi)\approx 0\). When the three class probabilities are comparable, the classification branch indicates uncertainty in the FOV state. In this case, the contribution of the segmentation map is naturally moderated by \(p_{\mathrm{I}}\), preventing the raw segmentation output from being over-interpreted as a reliable boundary when the model is uncertain whether the magnetopause is observable inside the FOV.

The final binary magnetopause-envelope mask is then obtained by thresholding the fused probability map:
\begin{equation}
\hat{M}(\theta,\phi)
=
\mathbb{I}\left[M(\theta,\phi)\ge \tau\right],
\label{eq:final_mask}
\end{equation}
where \(\tau=0.5\) is used in this study.

This dual-branch design explicitly accounts for the effects of  noise and observation geometry on magnetopause visibility. It enables stable magnetopause detection when the boundary is partially visible or lies outside the SXI FOV. For cases in which the magnetopause is fully contained within the FOV, the framework provides reliable pixel-level boundary information that can be readily combined with conventional inversion techniques for 3D magnetopause reconstruction.

\subsection{Loss Function}

During training, the network parameters are optimized by minimizing a loss function using the backpropagation algorithm \citep{backpropagation1986hinton,dl_loss_bp_2016goodfellow}. The total loss is defined as the weighted sum of a classification loss and a segmentation loss,
\begin{equation}
\mathcal{L}
=
\mathcal{L}_{\mathrm{cls}}
+
\alpha \, \mathcal{L}_{\mathrm{seg}},
\end{equation}
where $\alpha$ is a weighting coefficient that balances the contributions of the two tasks.

The classification loss $\mathcal{L}_{\mathrm{cls}}$ is formulated as a multiclass cross-entropy loss for the three-category magnetopause localization problem (Left, Inside, and Right),
\begin{equation}
\mathcal{L}_{\mathrm{cls}}
=
-\frac{1}{N}
\sum_{i=1}^{N}
\sum_{c \in \{L,I,R\}}
y_{i,c}
\log\left(p_{i,c}\right),
\end{equation}
where $y_{i,c}$ denotes the ground truth label of sample $i$ for class $c$, $p_{i,c}$ is the predicted class probability, and $N$ is the number of samples.

The segmentation loss $\mathcal{L}_{\mathrm{seg}}$ is defined as a weighted binary cross-entropy loss and is applied only when the magnetopause is predicted to be located within the SXI FOV. 
It is expressed as
\begin{equation}
\mathcal{L}_{\mathrm{seg}}
=
\frac{1}{M}
\sum_{j=1}^{M} p_{\mathrm{j,I}}
\left[
- y_j \log(\hat{y}_j)
-
(1-y_j)\log(1-\hat{y}_j)
\right],
\end{equation}
where $p_{\mathrm{j,I}}$ is the probability predicted by the classification head that the magnetopause lies inside the SXI FOV, $y_j$ and $\hat{y}_j$ denote the ground truth label and predicted probability of pixel $j$, respectively, and $M$ is the total number of pixels.

By modulating the segmentation loss with the classification probability $p_{\mathrm{I}}$, pixel-level supervision is emphasized primarily for samples in which the magnetopause is likely to be observable within the SXI FOV. For cases in which the magnetopause is located outside the FOV, the influence of the segmentation loss is naturally suppressed, thereby avoiding the imposition of physically meaningless boundary constraints. 
This coupling strategy improves the stability of joint optimization and enhances model robustness under complex observational conditions.

\section{Experiments and Metrics} \label{sec_expsAndmetrics}

\subsection{Training Configuration}

Model training is performed using the Adam optimizer \citep{Adam2015Kingma} with a learning rate of $1 \times 10^{-4}$. The maximum number of training epochs is set to 500, and the batch size is fixed at 800. To mitigate overfitting, an early stopping strategy is employed. 
Training is terminated if the validation loss does not decrease for 50 consecutive epochs. In the joint loss function, the weighting coefficient between the classification and segmentation losses is set to $\alpha = 1$, which assigns equal importance to the two tasks. Preliminary experiments indicate that this setting provides stable convergence and balanced performance for both magnetopause localization and boundary segmentation.


During training, horizontal flipping is applied as a data augmentation technique to enhance model adaptability to variations in observation geometry. The original simulated SXI images have a spatial size of $15 \times 25$ pix$^2$, which is not fully compatible with the multiple downsampling operations in the U-Net architecture. Therefore, bilinear interpolation is applied at the network input stage to resample the images to $32 \times 32$ pix$^2$. The predicted segmentation masks are subsequently resized back to the original size of $15 \times 25$ pix$^2$, using nearest-neighbor interpolation for pixel-wise comparison with the reference magnetopause envelope. This procedure ensures architectural compatibility while avoiding additional smoothing effects in the segmentation results.

\subsection{Evaluation Metrics}


The classification performance of magnetopause localization with respect to the SXI FOV is evaluated using a confusion matrix. 
For the three-class classification task (Left, Inside, and Right), the confusion matrix $\mathbf{M} \in \mathbb{R}^{N \times N}$ with $N = 3$ is defined such that the element $M_{ij}$ denotes the number of samples belonging to class $i$ that are predicted as class $j$.

Based on the confusion matrix, overall accuracy (Accuracy) and macro-averaged F1 score (Macro-F1) are adopted as evaluation metrics. 
The overall accuracy is defined as Equation~(\ref{equ_Acc}).
\begin{equation}
\mathrm{Accuracy}
=
\frac{\sum_{i=1}^{N} M_{ii}}{\sum_{i=1}^{N} \sum_{j=1}^{N} M_{ij}}.
\label{equ_Acc}
\end{equation}

For each class $k$, the precision and recall are defined as Equations~(\ref{equ_Pre_Recall}).
\begin{equation}
\mathrm{Precision}_k
=
\frac{M_{kk}}{\sum_{j=1}^{N} M_{jk}},
\qquad
\mathrm{Recall}_k
=
\frac{M_{kk}}{\sum_{j=1}^{N} M_{kj}}.
\label{equ_Pre_Recall}
\end{equation}

The F1 score for class $k$ is given by Equation~(\ref{equ_F1k}).
\begin{equation}
\mathrm{F1}_k
=
\frac{2 \cdot \mathrm{Precision}_k \cdot \mathrm{Recall}_k}
{\mathrm{Precision}_k + \mathrm{Recall}_k}.
\label{equ_F1k}
\end{equation}

The Macro-F1 score is computed as the arithmetic mean of the class-wise F1 scores.
\begin{equation}
\mathrm{Macro\text{-}F1}
=
\frac{1}{N}
\sum_{k=1}^{N}
\mathrm{F1}_k.
\end{equation}

In addition to overall accuracy and Macro-F1 score, Cohen’s kappa coefficient is used to quantify the agreement between the predicted and true magnetopause position classes while accounting for agreement expected by chance.
Cohen’s kappa is defined as Equation~(\ref{equ_kappa}).
\begin{equation}
\kappa = \frac{p_o - p_e}{1 - p_e},
\label{equ_kappa}
\end{equation}
where $p_o$ denotes the observed classification accuracy and $p_e$ represents the expected agreement under random classification.
The kappa coefficient provides a more conservative and robust assessment of classification performance, particularly for multi-class problems.


The segmentation performance of magnetopause boundary detection is quantified using pixel-level evaluation metrics. The predicted binary mask is compared with the reference LOS magnetopause envelope on a pixel-by-pixel basis. Each pixel is classified as true positive (TP), false positive (FP), true negative (TN), or false negative (FN), according to the relationship between the prediction and the ground truth.
Based on these definitions, pixel accuracy, precision, recall, and F1 score are computed as Equation~(13-15).
\begin{equation}
\mathrm{Pixel\ Accuracy}
=
\frac{\mathrm{TP} + \mathrm{TN}}
{\mathrm{TP} + \mathrm{FP} + \mathrm{TN} + \mathrm{FN}},
\end{equation}
\begin{equation}
\mathrm{Precision}
=
\frac{\mathrm{TP}}{\mathrm{TP} + \mathrm{FP}},
\qquad
\mathrm{Recall}
=
\frac{\mathrm{TP}}{\mathrm{TP} + \mathrm{FN}},
\end{equation}
\begin{equation}
\mathrm{F1}
=
\frac{2 \cdot \mathrm{Precision} \cdot \mathrm{Recall}}
{\mathrm{Precision} + \mathrm{Recall}}.
\end{equation}

In addition, the intersection over union (IoU) metric is employed to further assess the spatial consistency between the predicted and reference magnetopause envelopes. IoU is defined as the ratio of the intersection to the union of the predicted mask and the ground truth.
\begin{equation}
\mathrm{IoU}
=
\frac{\mathrm{TP}}
{\mathrm{TP} + \mathrm{FP} + \mathrm{FN}}.
\end{equation}

This metric is particularly suitable for evaluating connected boundary structures, as it directly reflects the overall overlap between the predicted magnetopause region and the reference envelope.

\subsection{Boundary-Location Metrics and Benchmark Protocols}

In addition to region-overlap metrics, we quantified magnetopause boundary localization on the native \(15\times25\) SXI angular grid. The fused \(32\times32\) probability map was bilinearly resampled to
the native grid and binarized at \(\tau=0.5\). The reference masks were restored to the native grid using nearest-neighbor interpolation. In the physical image convention, rows represent \(\phi\) and columns represent \(\theta\). The pixel-edge ranges were \(\theta\in[-7.5^\circ,7.5^\circ]\) and
\(\phi\in[-12.5^\circ,12.5^\circ]\), giving a pixel-center spacing of \(1^\circ\) along both axes.

For each \(\phi_j\) row, \(\hat{\theta}(\phi_j)\) and \(\theta^\ast(\phi_j)\) denote the predicted and reference boundary coordinates, respectively. Their mean row-wise absolute angular displacement was defined as

\begin{equation}
E_{\mathrm{row}}
=
\frac{1}{|\Omega|}
\sum_{j\in\Omega}
\left|
\hat{\theta}(\phi_j)-\theta^\ast(\phi_j)
\right|,
\label{eq:rowwise_boundary_error}
\end{equation}

where \(\Omega\) contains the rows in which both boundaries are defined. The central-row error was evaluated at the sampled latitude closest to \(\phi=0^\circ\):

Let \(\mathcal{J}=\{1,\ldots,N_{\phi}\}\) denote the set of row indices on the native SXI grid, and let \(\phi_j\) be the angular coordinate at the center of row \(j\). The image-center row is defined as

\begin{equation}
j_0 =
\operatorname*{arg\,min}_{j\in\mathcal{J}}
\left|\phi_j\right|.
\label{eq:central_row}
\end{equation}

Let \(\hat{\theta}_j\) and \(\theta_j^\ast\) denote the predicted and
reference magnetopause boundary coordinates, respectively, along row
\(j\). The central-row boundary error is then

\begin{equation}
E_0 =
\left|
\hat{\theta}_{j_0}
-
\theta_{j_0}^\ast
\right|.
\label{eq:central_boundary_error}
\end{equation}

Here, \(j_0\) is the row whose center is closest to
\(\phi=0^\circ\). For the native grid used in this study, this row is
centered at \(\phi=0^\circ\). The quantity \(E_0\), reported in degrees,
measures the absolute displacement of the predicted boundary from the
reference boundary in the \(\theta\) direction at the center of the
image. It is calculated only when both the predicted and reference
boundaries are defined at row \(j_0\); otherwise, the central-row error
for that image is treated as unavailable.

Boundary-distance metrics were evaluated for reference-Inside samples. As a complementary traditional-method case study, we used the publicly available synthetic SXI image analyzed by \citet{tfa2020}. The TFA curve was reconstructed from the reported best-fit parameters and compared with MP-DualNet and the row-wise intensity-maximum locus, \(\max(I_X)\), relative to the projected MHD-fit reference under the same viewing geometry.

\section{Results} \label{sec_results}
\subsection{Training Convergence and Model Optimization}


\begin{figure}[htbp]
\centering
\includegraphics[width=\linewidth]{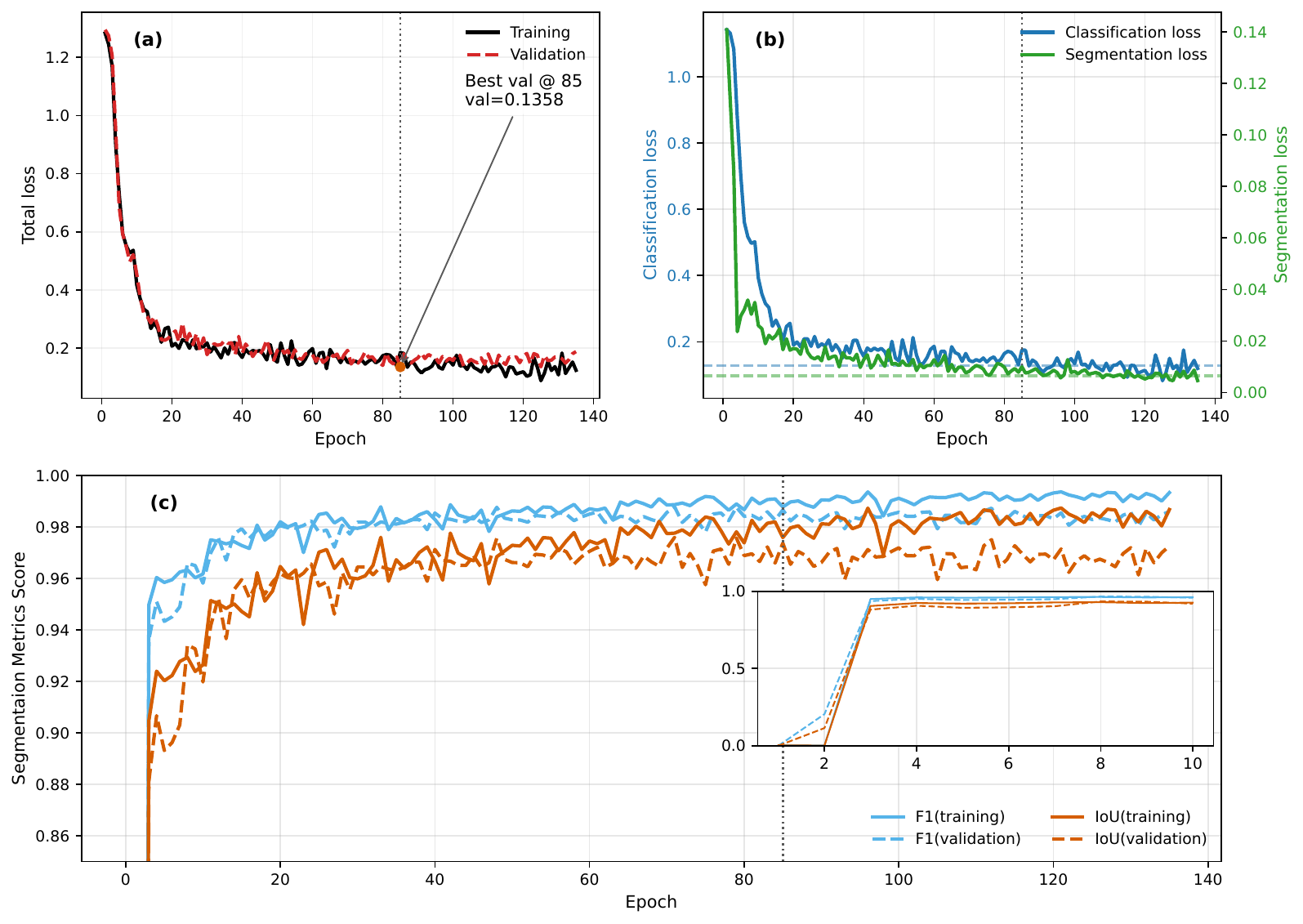}
\caption{
Training history of the MP-DualNet framework.
(a) Total training loss (black solid line) and validation loss (red dashed line) as functions of epoch. The gray dotted vertical line marks Epoch 85, at which the minimum validation loss of
0.1358 is obtained, and the orange marker indicates the corresponding checkpoint.
(b) Evolution of the classification-branch loss (blue line; left vertical axis) and segmentation-branch loss (green line; right vertical axis). The gray dotted vertical line again indicates the selected checkpoint at Epoch 85.
(c) F1 score (blue) and IoU (orange) for the training (solid lines) and validation (dashed lines) sets. The inset enlarges the first 10 epochs using the same color and line-style conventions, highlighting the rapid improvement in segmentation performance during the initial optimization stage.
}
\label{fig:loss_curve}
\end{figure}

Figure~\ref{fig:loss_curve} summarizes the training dynamics of the MP-DualNet framework. The evolution of the training and validation losses indicates rapid convergence during the early training stage, followed by a gradual stabilization of the optimization process. Figure~\ref{fig:loss_curve}(a) shows the evolution of the training and validation losses for the magnetopause localization task. During the initial training stage (approximately the first 20 epochs), the total loss decreases rapidly to about 0.2, then gradually converges to a stable minimum. While the training loss continues to decrease throughout the optimization process, the validation loss shows only marginal improvement beyond 85 epochs, suggesting the model has reached an optimal generalization state.

As shown in Figure~\ref{fig:loss_curve}(b), the loss curves associated with the classification and segmentation components exhibit consistent convergence behavior. Both losses decrease rapidly during the early training phase and stabilize after approximately 85 epochs, suggesting balanced joint optimization of the two tasks. 
Figure~\ref{fig:loss_curve}(c) presents the evolution of the F1 score and IoU for the training and validation sets. The inset enlarges the first 10 epochs and displays the same four metric curves as the main panel. It shows that the largest performance gains occur during the first few epochs: the F1 and IoU scores increase sharply during this period and reach approximately 0.9 or higher by Epoch 4. After this initial increase, the metrics improve more gradually. At the selected checkpoint, the F1 score exceeds 0.98 and the IoU exceeds 0.97 for both the training and validation sets. The small remaining difference between the training and validation curves indicates similar performance on the two subsets under the adopted data-partitioning strategy.

To mitigate overfitting, early stopping is applied, and training is terminated at epoch 135. The model corresponding to the minimum validation loss is selected for subsequent analysis. Although this model does not achieve the lowest training loss, it provides the best balance between accuracy and generalization capability.

\subsection{Coarse Localization of the Magnetopause}

\begin{figure}[htbp]
\centering
\includegraphics[width=\linewidth]{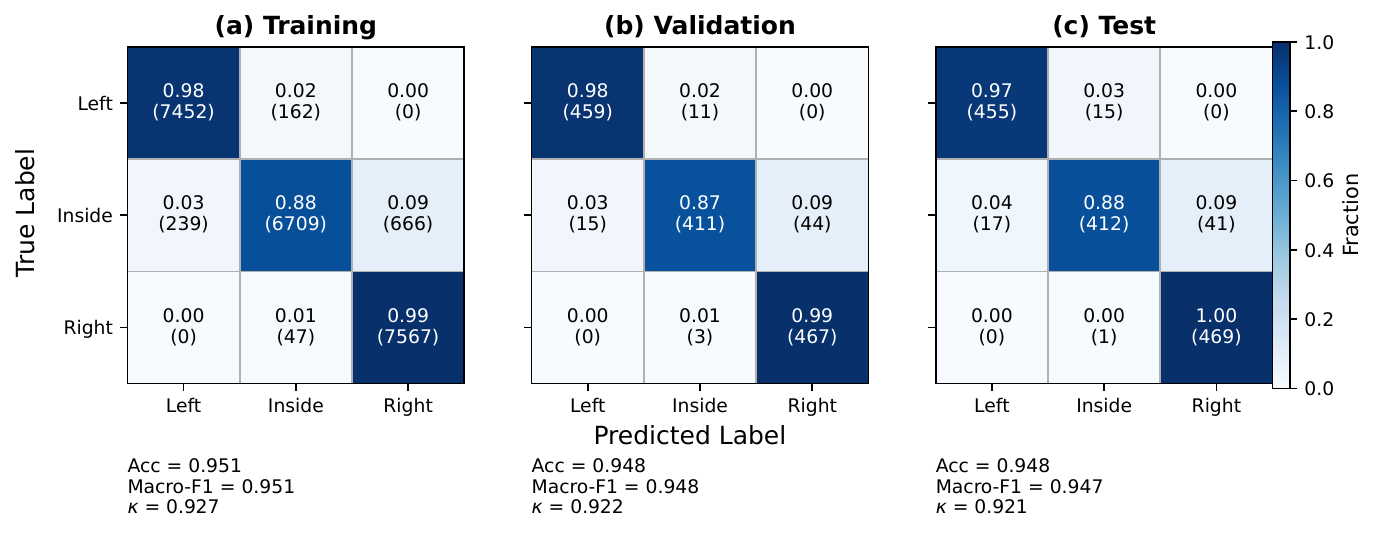}
\caption{Confusion matrices for the magnetopause position classification on the (a) training, (b) validation, and (c) test datasets. 
Each matrix shows the normalized fraction of samples for three classes: Left, Inside, and Right, which represent the relative position of the magnetopause with respect to the SXI FOV. Rows correspond to the true labels, and columns correspond to the predicted labels. The numbers in each cell indicate the normalized fraction, with the absolute sample counts given in parentheses. The quantitative performance metrics, including overall accuracy (Acc), macro-averaged F1 score (Macro-F1), and Cohen’s kappa coefficient ($\kappa$), are reported below each panel, demonstrating consistent classification performance and good generalization capability of the model.
}
\label{fig:confusion_matrix}
\end{figure}


Figure~\ref{fig:confusion_matrix} presents the confusion matrices of the coarse localization model evaluated on the training, validation, and test datasets. The quantitative results demonstrate consistent and stable classification performance across all datasets. Specifically, the model achieves an accuracy of 0.951, a Macro-F1 score of 0.951, and a Cohen’s kappa coefficient of 0.927 on the training set. On the validation set, the accuracy, Macro-F1 score, and kappa coefficient reach 0.945, 0.948, and 0.922, respectively. For the independent test set, the corresponding values are 0.948, 0.947, and 0.921.

The confusion matrices indicate that the model achieves high classification accuracy for the Left and Right categories, corresponding to cases in which the magnetopause lies outside the SXI FOV. Misclassifications are primarily associated with the Inside category, where some samples are incorrectly classified as Right. These cases typically correspond to scenarios in which the magnetopause is located near the boundary of the FOV, resulting in incomplete envelope signatures and reduced discriminative information in the SXI images.

Overall, the coarse localization model provides reliable identification of the magnetopause position relative to the SXI FOV, thereby supplying a robust prior for subsequent fine-scale boundary detection.

\subsection{Fine-Scale Magnetopause Boundary Localization}

\begin{table}[t]
\centering
\caption{Segmentation performance of MP-DualNet on the training, validation, and test sets. All metrics are computed using the final trained model.}
\label{tab:perf_main}
\begin{tabular}{lccccc}
\hline
Dataset & Accuracy & Precision & Recall & F1-score & IoU \\
\hline
Training   & 0.991 & 0.993 & 0.986 & 0.990 & 0.979 \\
Validation & 0.987 & 0.986 & 0.983 & 0.984 & 0.969 \\
Test       & 0.988 & 0.991 & 0.981 & 0.986 & 0.972 \\
\hline
\end{tabular}
\end{table}

For samples classified by the coarse localization model as Inside, fine-scale boundary localization is subsequently performed using the segmentation branch. The segmentation network produces a continuous pixel-wise probability map representing the likelihood that each pixel belongs to the magnetospheric side of the magnetopause. The segmentation output is combined with the classification probability and binarized using a fixed threshold ($\mathrm{Th} = 0.5$) to obtain the final magnetopause envelope mask.

Table~\ref{tab:perf_main} summarizes the segmentation performance on the training, validation, and test datasets. Consistent performance across the three datasets indicates good generalization capability. On the independent test set, the model achieves a segmentation accuracy of 0.988, a precision of 0.991, a recall of 0.981, an F1 score of 0.986, and an IoU of 0.972. These results demonstrate that MP-DualNet can reliably and accurately identify the magnetopause region under diverse observational conditions. To quantify sampling variability in the test results, we performed an image-level nonparametric bootstrap analysis (Table~\ref{tab:bootstrap_uncertainty}). For each of 5,000
replicates, 1,515 images were sampled with replacement from the balanced test set, and the classification and segmentation metrics were
recalculated. The 95\% confidence intervals were defined by the 2.5th and 97.5th percentiles of the bootstrap distributions. Before evaluating the segmentation metrics, the continuous model output was mapped back to the native \(15\times25\) grid and threshold at \(\tau=0.5\). The relatively narrow intervals indicate limited variation under image-level resampling of the present test set. These intervals describe sampling uncertainty for one fixed trained model.

\begin{table}[htbp]
\centering
\caption{Performance and image-level bootstrap uncertainty on the full balanced test set. Fused-mask metrics were evaluated on the native
\(15\times25\) SXI grid. The intervals represent 95\% percentile bootstrap confidence intervals from 5,000 resamples.}
\label{tab:bootstrap_uncertainty}
\begin{tabular}{lcc}
\hline
Metric & Estimate & 95\% confidence interval \\
\hline
Classification accuracy & 0.948 & 0.936--0.959 \\
Classification Macro-F1 & 0.947 & 0.935--0.958 \\
Classification kappa    & 0.921 & 0.903--0.938 \\
Fused-mask accuracy     & 0.988 & 0.985--0.991 \\
Fused-mask precision    & 0.992 & 0.988--0.995 \\
Fused-mask recall       & 0.981 & 0.974--0.986 \\
Fused-mask F1           & 0.986 & 0.982--0.989 \\
Fused-mask IoU          & 0.972 & 0.965--0.979 \\
\hline
\end{tabular}
\end{table}

\begin{figure}[htbp]
\centering
\includegraphics[width=\linewidth]{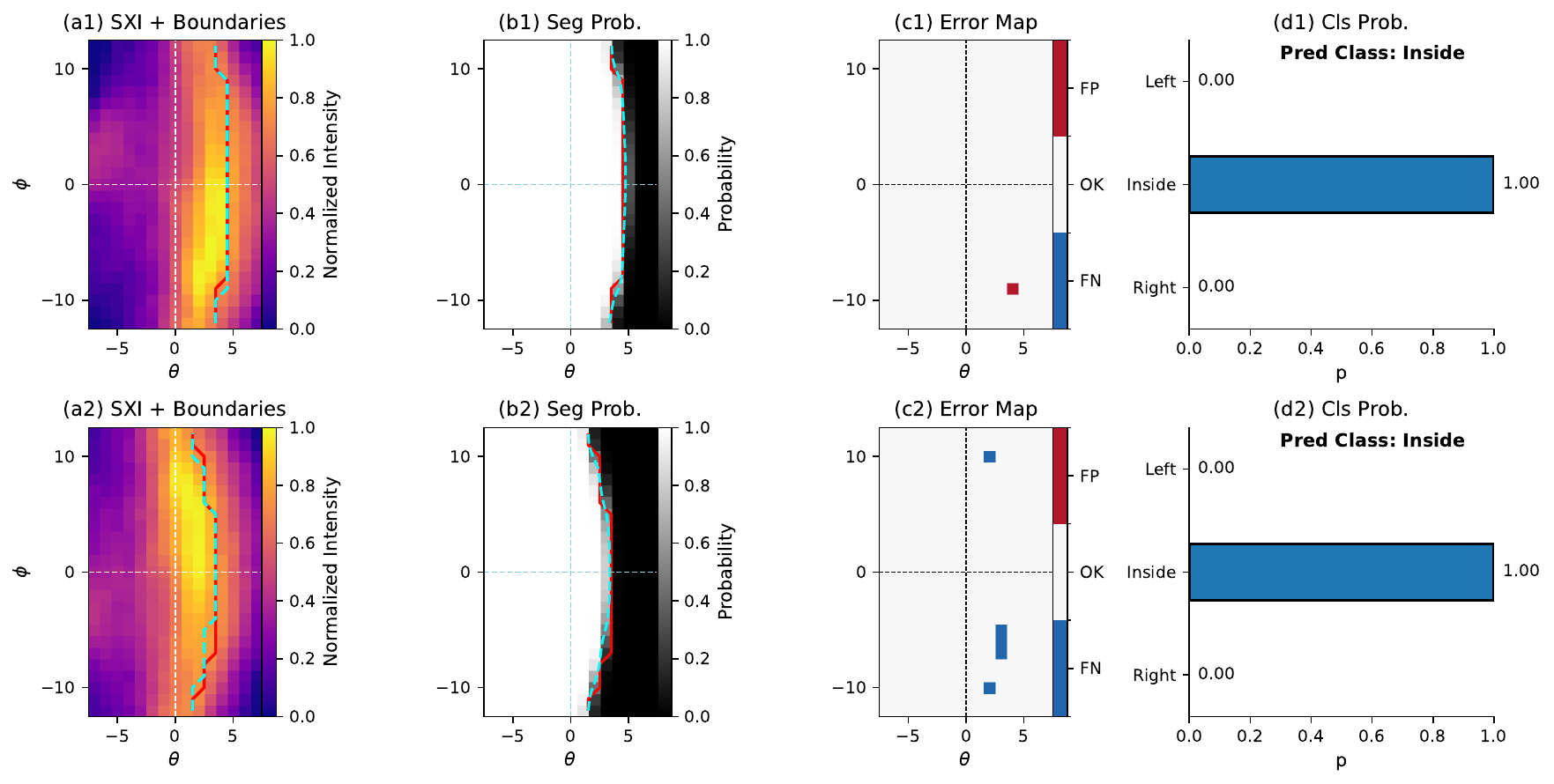}
\caption{
Representative examples of magnetopause boundary detection results for two test samples classified as Inside.
Each row corresponds to one representative case.
Panels (a1) and (a2) show the normalized LOS-integrated SXI images overlaid with the reference magnetopause envelope (red solid line) and the predicted boundary (cyan dashed line).
Panels (b1) and (b2) present the pixel-wise segmentation probability maps.
Panels (c1) and (c2) show the spatial distribution of segmentation errors, including false positives (FP) and false negatives (FN).
Panels (d1) and (d2) display the probability outputs of the classification branch for the three categories: Left, Inside, and Right, corresponding to the relative position of the magnetopause with respect to the SXI FOV.
}
\label{fig:clean_result}
\end{figure}

Figure~\ref{fig:clean_result} presents two representative samples for which the magnetopause envelope is fully contained within the SXI FOV. 
As shown in panels (a1) and (a2) of Figure~\ref{fig:clean_result}, enhanced soft X-ray emission regions are present in the SXI images, while the local intensity maxima do not always coincide with the magnetopause envelope derived from the hybrid simulations. 
This discrepancy reflects the fact that the soft X-ray intensity distribution is not uniquely determined by boundary geometry and is influenced by LOS integration. Despite this complexity, the model assigns high confidence to the Inside category for both samples, as shown in panels (d1) and (d2) of Figure~\ref{fig:clean_result}, and successfully recovers the overall shape and location of the magnetopause envelope.
The predicted boundaries closely follow the reference envelopes over most azimuthal angles, with only limited false detections or omissions in localized regions, as illustrated in panels (c1) and (c2) of Figure~\ref{fig:clean_result}. Correspondingly, the IoU and F1 scores remain high, indicating good spatial agreement between the predicted and reference magnetopause envelopes. 


\subsection{Robustness under Different Noise Levels}

To evaluate the robustness of the proposed framework under varying observational noise conditions, a representative event identical to that shown in Figure~\ref{fig:clean_result}(a) is selected for analysis. The reference magnetopause envelope is kept fixed, while  noise with intensity levels ranging from 5 to 50~keV\,cm$^{-2}$\,s$^{-1}$\,sr$^{-1}$, in increments of 5~keV\,cm$^{-2}$\,s$^{-1}$\,sr$^{-1}$, is superimposed on the corresponding SXI LOS-integrated images. The noise-contaminated images are then processed by the trained model to assess its stability under increasing noise levels.

\begin{figure}[htbp]
    \centering
    \includegraphics[width=\linewidth]{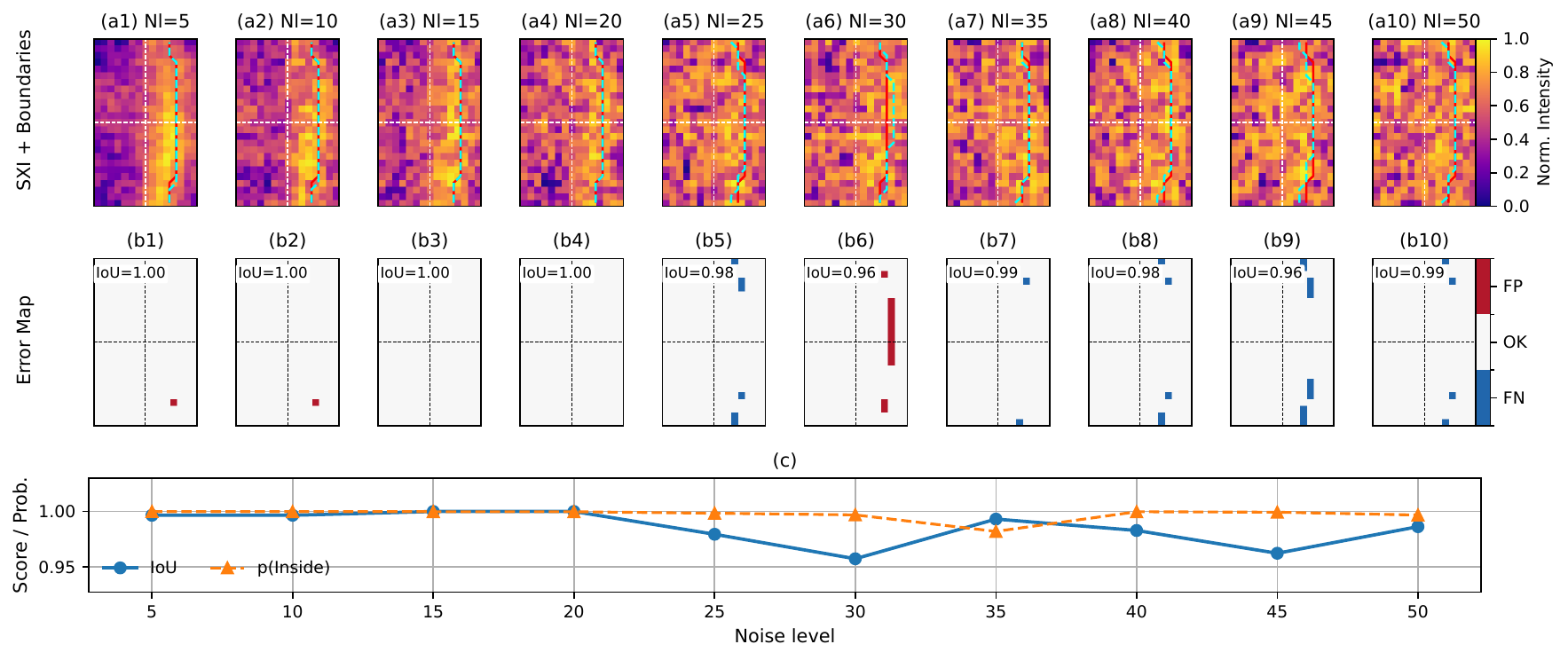}
    \caption{Magnetopause boundary detection results under different  noise levels.
    Top row: normalized SXI images overlaid with the reference (red solid line) and predicted (cyan dashed line) magnetopause envelopes.
    Middle row: spatial distribution of false positives (FP) and false negatives (FN) relative to the reference envelope.
    Bottom row: variation of the intersection over union (IoU) and the classification probability $p(\mathrm{Inside})$ as a function of noise level.}
    \label{fig:Noise_level}
\end{figure}

Figure~\ref{fig:Noise_level} provides an overview of the model predictions under different noise levels. Across the entire noise range considered, the model maintains effective constraints on the large-scale position and overall morphology of the magnetopause. The predicted envelopes remain largely consistent with the reference boundary, with discrepancies primarily appearing as localized and spatially isolated false positives or false negatives near the boundary region. These errors do not lead to systematic shifts of the magnetopause location.

The quantitative results further indicate that the IoU remains at a consistently high level, ranging approximately from 0.96 to 1.00 across all noise levels. At the same time, the classification branch assigns a high confidence to the Inside category, with $p_\mathrm{I}$ remaining close to unity and exhibiting only minor fluctuations at a few noise levels. Notably, neither the IoU nor $p_\mathrm{I}$ exhibits a monotonic degradation with increasing noise intensity. This behavior suggests that the proposed framework is resilient to random  noise under the experimental conditions considered. Taken together, the visual inspection and quantitative metrics demonstrate that the model achieves stable and consistent magnetopause boundary extraction over a broad range of noise levels.

\subsection{Extra Validation Using a Storm Event}

To assess the generalization capability of MP-DualNet under more complex space weather conditions, an independent storm event that was not included in the training/validation/test sets is selected for extra validation. Specifically, global hybrid simulation results for the May 2024 geomagnetic storm presented by \citet{2024storm_mpcusps_sxi_2025ng} are employed to generate a sequence of synthetic SXI LOS-integrated images. These data are used solely as external test inputs and were not used for model training or validation.

\begin{figure}[htbp]
    \centering
    \includegraphics[width=\linewidth]{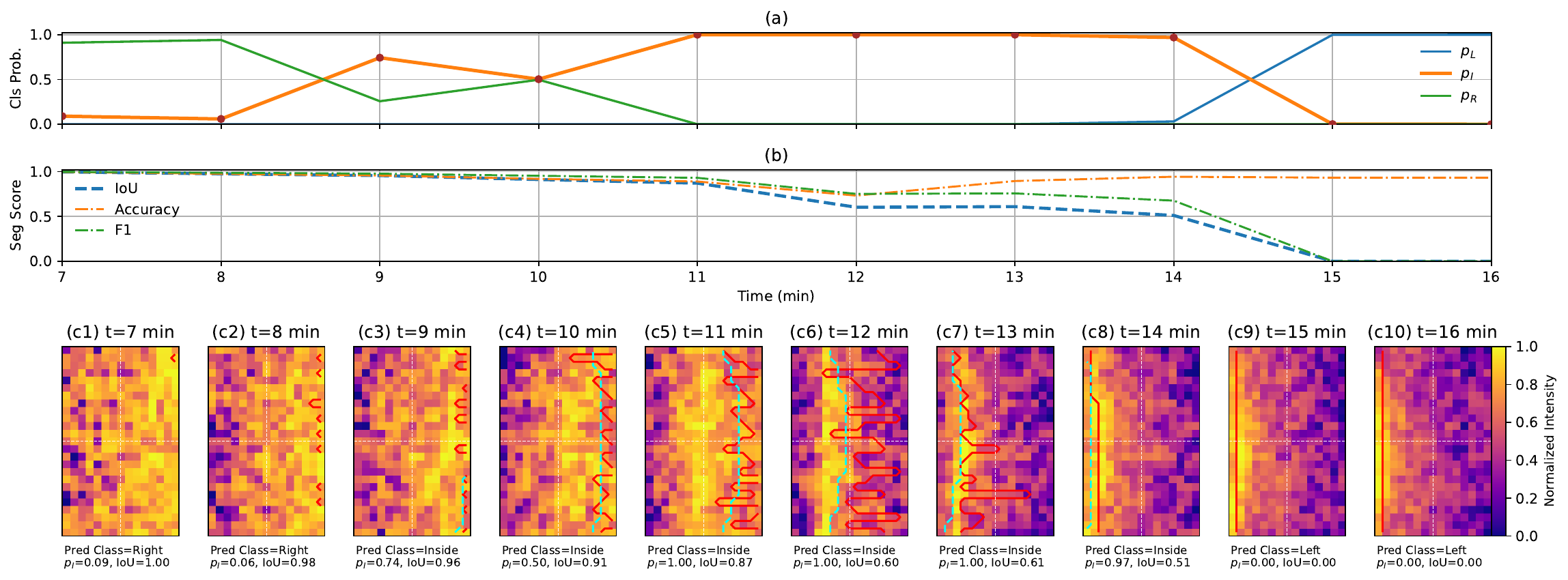}
    \caption{External validation of MP-DualNet during a storm-time event.
    Top panels: temporal evolution of classification probabilities for the Left, Inside, and Right categories.
    Middle panels: variation of segmentation metrics, including IoU, Accuracy, and F1 over time.
    Bottom panels: selected SXI frames overlaid with the reference magnetopause envelope (red solid line) and the predicted boundary (cyan dashed line).}
    \label{fig:storm_event}
\end{figure}

Figure~\ref{fig:storm_event} presents an overview of the model predictions over a selected time interval of the storm event. The classification branch successfully captures the large-scale geometric evolution of the magnetopause relative to the SXI FOV. At early times, the predicted category is dominated by the Right class, with a high $p_\mathrm{R}$, indicating that the magnetopause lies near the edge of the FOV. As the event evolves, the dominant prediction transitions to the Inside category, with $p_\mathrm{I}$ approaching unity, corresponding to the phase during which the magnetopause moves into and remains within the SXI FOV.

At the same time, the predicted magnetopause boundaries show good agreement with the reference envelopes in terms of large-scale position and overall morphology across multiple frames. The model reproduces the primary temporal evolution of the magnetopause during the storm, demonstrating its ability to generalize to an independent event characterized by stronger solar wind driving and enhanced magnetospheric dynamics.

It is noted that the segmentation performance, particularly the IoU metric, degrades noticeably at several time steps. Visual inspection indicates that, during these intervals, the reference magnetopause envelope displays increased small-scale complexity, such as stronger local fluctuation and non-smooth structures. Given the spatial resolution of the input SXI images, the presence of  noise, and the fact that the model is primarily trained to capture large-scale envelope morphology, deviations in fine-scale boundary details are expected.

Overall, this external case study demonstrates that MP-DualNet is capable of capturing the global evolution of the magnetopause under storm conditions, supporting its transferability to independent and dynamically complex scenarios. At the same time, the results highlight remaining challenges in resolving fine-scale boundary structures during periods of strong disturbance.

\subsection{Physical Boundary Accuracy and Benchmark Comparisons}

Among the test set samples, the mean absolute angular displacement was \(0.488^\circ\), while the mean central-row absolute angular error was \(0.4112^\circ\). These physical metrics quantify the boundary localization and complement the IoU and F1 scores.

\begin{figure}[htbp]
    \centering
    \includegraphics[width=\textwidth]
    {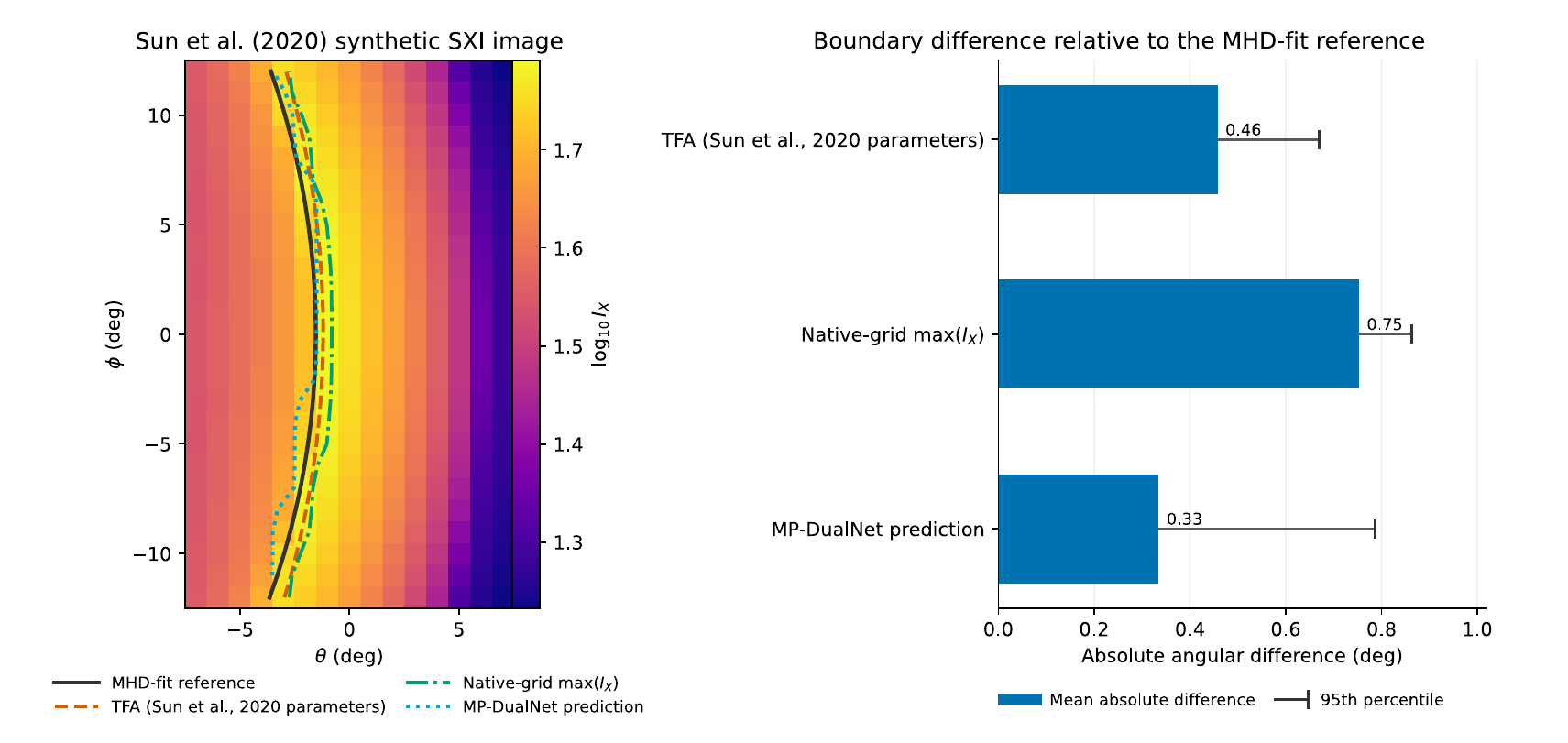}
    \caption{
    Comparison of magnetopause boundary estimates for the public 
    synthetic SXI case analyzed by \cite{tfa2020}.
    (a) Synthetic SXI image in the \((\theta,\phi)\) angular
    coordinates, overlaid with the projected MHD-fit reference
    (black solid), the TFA boundary reconstructed using the parameters
    reported by \cite{tfa2020} (\(r_0=8.0\,R_E\), \(\alpha_y=0.8\), and \(\alpha_z=0.2\);
    orange dashed), the row-wise intensity-maximum locus
    \(\max(I_X)\) (green dash-dotted), and the MP-DualNet prediction
    (cyan dotted).
    (b) Absolute angular differences from the projected MHD-fit
    reference. Horizontal bars represent the mean absolute difference
    over the rows in which both the estimated and reference boundaries
    are defined. The terminal markers indicate the corresponding
    95th percentiles.
    }
    \label{fig:MPdualnet_tfa}
\end{figure}

Figure~\ref{fig:MPdualnet_tfa} provides a complementary comparison with a traditional reconstruction method using the public synthetic SXI case analyzed by \citet{tfa2020}. Relative to the projected
MHD-fit reference, the mean absolute angular difference was \(0.33^\circ\) for MP-DualNet, \(0.46^\circ\) for the TFA boundary reconstructed from the parameters reported by \citet{tfa2020}, and
\(0.75^\circ\) for the native-grid \(\max(I_X)\) locus. MP-DualNet therefore produced the smallest mean difference in this example. However, the 95th-percentile comparison shows that the smaller mean
does not imply uniformly smaller local deviations along the entire boundary. Because this analysis concerns a single external case and uses a TFA curve reconstructed from published parameters, it is
presented as a case-specific consistency test rather than evidence of dataset-wide superiority over TFA.

\section{Discussions and Conclusions} \label{sec_discussion_conclusion}

\subsection{Generalization Across Different Solar Wind Conditions}

The SXI images used in this study are generated from global hybrid simulations under three distinct sets of upstream solar wind and IMF conditions. Additional  noise levels are introduced to further diversify the dataset, resulting in substantial variations in the pixel intensity distributions across different subsets. Under such conditions, a global min-max normalization applied uniformly across the entire dataset may compress physically meaningful contrast and weaken boundary-related features near the magnetopause.

To address this issue, an image-wise normalization strategy based on the 1st and 99th percentile pixel values is adopted during preprocessing. This approach suppresses extreme outliers while enhancing intensity gradients across the magnetopause boundary. As a result, the network is guided to focus on relative intensity variations rather than absolute intensity levels. The experimental results demonstrate that this strategy improves model stability across different solar wind conditions. Even under low solar wind density cases (e.g., $N_{\mathrm{sw}} \approx 7~\mathrm{cm^{-3}}$), the model remains capable of identifying the large-scale location of the magnetopause.

These findings indicate that MP-DualNet does not rely on intensity patterns specific to individual solar wind conditions. Instead, the network learns features associated with the geometric configuration and spatial structure of the magnetopause. Such generalization capability is essential for practical applications of SXI-based magnetopause detection, particularly for the SMILE mission, where solar wind conditions are expected to vary significantly over time.

\subsection{Magnetopause Detection Under Limited FOV Conditions}

Unlike traditional magnetopause reconstruction methods, MP-DualNet is explicitly designed to account for cases in which the magnetopause is not fully contained within the SXI FOV \citep{fov_cls_mp_rf_2023wang}. During SMILE operations, variations in spacecraft orbit and pointing geometry may cause the magnetopause to be partially visible or entirely outside the observable region, rendering many conventional approaches inapplicable.

The coarse localization module introduced in MP-DualNet enables the model to infer the relative position of the magnetopause with respect to the SXI FOV under incomplete observational conditions. This capability provides a physically meaningful prior for subsequent processing. The results show that, when only a limited portion of the magnetopause envelope is visible near the FOV boundary, classification errors mainly arise from confusion between the Inside and Left/Right categories. This behavior reflects the intrinsic difficulty of geometric inference when boundary information is incomplete.

For cases in which the magnetopause is fully contained within the FOV, the classification results are notably reliable. This property supports a hybrid strategy that combines deep learning and traditional inversion techniques. In such a framework, MP-DualNet can be used to determine magnetopause visibility and provide coarse localization, while physics-based methods, such as the tangential fitting approach (TFA), can be selectively applied for 3D reconstruction. This combination has the potential to reduce misinterpretation and improve the robustness of magnetopause reconstruction under realistic observation constraints.

\subsection{Robustness to Noise and Observational Uncertainty}

In realistic space environments, SXI observations are inevitably affected by instrumental noise and diffuse cosmic X-ray background emission. Robust performance under such conditions is therefore a prerequisite for applying data-driven methods to space weather studies.

Both training and evaluation in this work explicitly incorporate SXI images with a wide range of  noise levels. The results demonstrate that MP-DualNet consistently recovers the large-scale position and overall morphology of the magnetopause from low to high noise conditions. Notably, the segmentation performance does not exhibit a monotonic degradation with increasing noise intensity. This behavior suggests that magnetopause identification by the proposed model relies primarily on global structural and gradient-related features rather than on local pixel-level intensity fluctuations.

The observed robustness arises from several contributing factors. First, the training dataset includes diverse noise realizations, exposing the network to a broad spectrum of signal-to-noise scenarios and reducing sensitivity to specific noise patterns. 
Second, the image-wise normalization strategy based on percentile clipping effectively suppresses extreme noise outliers while preserving contrast across the magnetopause boundary. Third, the relatively compact network architecture, together with early stopping during training, limits model capacity and mitigates overfitting to noise-dominated features.

Taken together, these factors enable MP-DualNet to maintain stable performance in the presence of significant observational uncertainty. This robustness holds significant implications for future soft X-ray observations, where noises are expected to vary considerably across different space environments.

\subsection{Storm-Time Magnetopause Dynamics: Case Study}

Although MP-DualNet demonstrates stable performance in capturing the large-scale geometric evolution of the magnetopause during the independent storm event, noticeable fluctuations in segmentation metrics are observed at several time steps. These variations indicate intrinsic challenges associated with magnetopause boundary detection under strongly disturbed conditions.

First, SXI images represent LOS–integrated soft X-ray emission. The spatial distribution of enhanced emission is jointly influenced by magnetosheath plasma density structures, viewing geometry, and background noise. As a result, intensity gradients do not necessarily form clear or persistent edges at the magnetopause location. During storm intervals, the magnetosheath becomes more structured and dynamic, and emission enhancements may appear more diffuse or exhibit reduced local contrast. Under such conditions, the amount of discriminative boundary information available in a single frame image is diminished, increasing the likelihood of localized false detections or omissions and leading to degraded pixel-level overlap metrics, such as IoU.

Second, uncertainty in the reference magnetopause envelope also contributes to the observed metric variability. In this study, the reference boundary is derived from global hybrid simulations using the mixing-rate criterion. During periods of strong disturbance, the extracted envelope may exhibit pronounced local undulations or non-smooth features. Because the network is primarily trained using envelope-level supervision, it tends to learn reproducible, large-scale geometry rather than local details. In addition, pixel-level metrics are highly sensitive to small spatial discrepancies. Consequently, localized deviations between the predicted and reference boundaries may result in substantial reductions in IoU, even when the predicted envelope remains visually consistent with the overall magnetopause configuration. This interpretation also explains the lack of a one-to-one correspondence between segmentation metrics and classification confidence during the storm event. High values of $p_\mathrm{I}$ indicate that the large-scale geometric state is correctly identified, whereas local boundary complexity and reference uncertainty primarily affect pixel-level overlap measures.

A comparison with the controlled noise experiments further suggests that the observed performance degradation is not governed solely by noise intensity. Instead, it is strongly influenced by the instantaneous detectability of the boundary, including contrast strength, gradient coherence, and geometric complexity. While MP-DualNet exhibits strong robustness to random background noise, storm conditions introduce additional challenges associated with rapidly evolving structures and complex LOS effects.

Overall, the storm case study confirms the transferability of MP-DualNet in capturing the global evolution of the magnetopause in independent and dynamically complex events. At the same time, it highlights key aspects that require further improvement to enhance reliability under strong disturbances.

\subsection{Relationship to Existing Fitting and Inversion Methods}

MP-DualNet, traditional fitting methods, and statistical inversion methods extract different information from SXI observations. MP-DualNet operates directly on an individual SXI image and provides two outputs: the position of the magnetopause relative to the SXI FOV and its
two-dimensional projected envelope when the boundary is visible. In contrast, the tangent fitting approach (TFA) estimates magnetopause geometry by fitting a prescribed boundary model to the tangent signature identified in an SXI image.

We compared MP-DualNet with TFA using the public synthetic SXI image analyzed by \citet{tfa2020}. All boundary differences were calculated relative to the same projected MHD-fit reference. The mean absolute angular difference was \(0.33^\circ\) for MP-DualNet, compared with \(0.46^\circ\) for the TFA curve reconstructed from the parameters reported by \citet{tfa2020} and \(0.75^\circ\) for the native-grid maximum-intensity locus. This common-image comparison shows that MP-DualNet recovers a projected boundary more consistent with the reference geometry in this example. Because the comparison is based on a single public image, it should not be interpreted as evidence that MP-DualNet is generally more accurate than TFA.

Statistical inversion addresses a related but distinct physical problem. Using the viewing geometry and a prior model of the magnetosheath emissivity, \citet{mploc_statisticalinv_sxi_2024cucho} applied maximum a posteriori inversion to estimate the 3D
soft X-ray emissivity distribution and infer the magnetopause location along the Sun--Earth line. MP-DualNet does not perform this 3D inversion and does not directly estimate a subsolar standoff distance. Instead, it identifies whether a projected boundary is observable and determines its location in the SXI image plane. The two approaches could therefore provide complementary information. MP-DualNet provides an image-based visibility classification and a projected boundary estimate, whereas statistical inversion estimates the 3D emissivity distribution and magnetopause location under an explicit viewing geometry and prior model.

\subsection{Limitations and Future Applications}

The reference magnetopause boundaries used for training and evaluation are derived from global hybrid simulations using the mixing-rate criterion. They are represented as two-dimensional projected envelopes that describe the large-scale magnetopause geometry. The physical magnetopause, however, is a finite-thickness boundary layer containing plasma structures over a range of spatial and temporal scales. The outputs of MP-DualNet should therefore be interpreted as estimates of the large-scale projected envelope, rather than as reconstructions of the internal structure or detailed plasma physics of the boundary layer.

The range of physical conditions represented in the training data is also limited. The three global hybrid simulation runs use fixed upstream solar-wind and IMF conditions. The grouped data partitioning strategy ensures that the noise realizations derived from a given source image remain in the same subset, thereby preventing closely related images from being shared between training and evaluation. The resulting test performance therefore measures transfer to unseen source-image groups within the three prescribed upstream regimes. It does not by itself demonstrate generalization to previously unseen solar-wind conditions or rapidly varying upstream forcing. Two external evaluations provide additional evidence beyond this grouped test. The independent storm sequence examines transfer to a time-dependent magnetopause configuration, while the public synthetic SXI image analyzed by \citet{tfa2020} examines transfer to an image generated independently of the present training dataset. MP-DualNet captured the large-scale magnetopause evolution in the storm sequence and recovered a boundary consistent with the projected MHD-fit reference in the \citet{tfa2020} case. These results indicate that the model is not limited to reproducing the noise realizations or source images used during training. Nevertheless, additional simulations covering a broader range of dynamic pressure, IMF orientation, and time-dependent upstream conditions are required, and the performance on SMILE observations is to be tested.

Under strongly disturbed conditions, the magnetopause may exhibit pronounced multi-scale structures and rapid temporal variability, including Kelvin–Helmholtz instabilities and local deformations driven by magnetosheath high-speed jets (HSJs) \citep{Guo2010KH_MHD,Sun2014KH_SXI,PfauKempf2020MagnetopauseReconnection,fatemi2024magnetosheathjets,globalsi_sxi_deformations_2023yang}. 

Although previous studies suggest that such features may be detectable in soft X-ray imaging, their characteristic spatial and temporal scales are often below the resolution and signal-to-noise regime of the data used in this work. Consequently, MP-DualNet is currently more effective at recovering the dominant large-scale arc-shaped envelope than at resolving local perturbations. 
This limitation likely arises from a combination of factors, including the relatively low occurrence rate of small-scale events in the training data, the intrinsic resolution and noise constraints of simulated SXI images, and the limited sensitivity of pixel-overlap–based loss functions and metrics to fine boundary details.

Several directions can be pursued to extend the present framework. First, refined supervision and evaluation can be achieved by replacing hard binary boundary labels with boundary bands or probabilistic references that reflect the magnetopause's finite thickness and uncertainty. Complementary geometric metrics, such as mean boundary distance, could be incorporated alongside Accuracy and IoU score to provide a more comprehensive assessment of boundary localization errors. Second, multiscale and local feature enhancement strategies could be introduced to improve sensitivity to localized curvature variations and asymmetric structures. 
Third, temporal information exploitation represents a natural extension, as the current model operates on a single-frame SXI image and does not explicitly enforce temporal consistency \citep{temporalevolutionofmp_2025gu}. Incorporating a temporal model or inter-frame coherence constraints may enhance stability and dynamic characterization during strongly disturbed periods. Finally, transfer to real observations will require domain adaptation and cross-validation using noise models, instrument responses, and, ultimately, in-orbit SMILE/SXI data.

Despite these limitations, MP-DualNet provides a robust end-to-end framework for magnetopause detection from soft X-ray images. Its stable, large-scale localization performance under diverse noise levels and viewing geometries demonstrates its potential as a practical tool for tracking magnetopause evolution and supporting subsequent 3D reconstruction in the context of the SMILE mission.

\subsection{Conclusions}

The magnetopause is a fundamental boundary formed by the interaction between the solar wind and the terrestrial magnetosphere. 
Its location and morphology provide direct insight into the response of the Sun-Earth system to upstream solar wind conditions. 
The SMILE mission, equipped with the SXI, will enable global remote sensing of the magnetopause. 
However, under dynamically varying solar wind conditions, especially during geomagnetic storms, the magnetopause can undergo rapid deformation and displacement, posing significant challenges for reliable detection and reconstruction.

In this study, we propose MP-DualNet, an end-to-end dual-branch deep learning framework for robust magnetopause detection and localization from simulated SXI images. By jointly modeling coarse localization through the classification branch and fine-scale boundary extraction through the pixel-level segmentation branch, the proposed method can handle a wide range of observation scenarios, including cases in which the magnetopause is fully contained within the FOV, partially visible, or entirely outside the observable region. 
The experimental results demonstrate stable and consistent performance across the training, validation, and test datasets, with classification accuracy exceeding 0.94 and segmentation IoU values above 0.97, indicating strong generalization capability.

By systematically incorporating multiple noise levels during data construction and training, together with a percentile-based normalization applied independently to each SXI image, the robustness of MP-DualNet under varying observational noise conditions is further confirmed. Application of the trained model to an independent storm event in May 2024 shows that MP-DualNet is capable of continuously tracking the large-scale evolution of the magnetopause under complex space weather conditions, with predicted boundaries remaining consistent with reference envelopes derived from global hybrid simulation at the macroscopic level.


Overall, MP-DualNet provides a robust framework for automated magnetopause detection from SXI images. By combining visibility classification with pixel-level segmentation, the method can handle cases where the magnetopause is partially visible or outside the SXI field of view. Its stable performance under varying noise levels and viewing geometries suggests good potential for supporting future magnetopause tracking and 3D reconstruction based on SXI observations.


\section*{Open Research}
The processed synthetic SXI datasets, pretrained models, and Python code for inference, evaluation, and visualization are archived at Zenodo \citep{zhong2026mpdualnet_archive}. The archive includes model architectures and trained weights, dependency requirements, a user guide, and executable examples for the test dataset, the Sun et al. (2020) comparison case, and the storm sequence. The README and notebook document input preprocessing, model loading, prediction, metric calculation, and figure generation, including the model checkpoints and postprocessing settings used in each example. This release supports use and evaluation of the supplied pretrained models.

The SXI data were generated from global hybrid simulations conducted by Zhongwei Yang and Tianran Sun. The deep learning implementation was adapted from the shallow U-Net framework described by \citet{liu2025shallowunet}, with task-specific modifications for magnetopause detection.


\section*{Conflict of Interest disclosure}
The authors declare there are no conflicts of interest for this manuscript.

\acknowledgments
This research is supported by the National Natural Science Foundation of China (NSFC 42521007, 42188101, 42150105, 42274210) and the National Key R\&D Program of China (2025YFF0512102). We also acknowledge the support from China's Space Origins Exploration Program, and the specialized Research Fund for State Key Laboratories of Solar Activity and Space Weather.
\bibliography{agu_revised}

\end{document}